\documentclass[11pt,a4paper,dvipsnames]{report}      

\usepackage{warwickthesis,setspace,graphicx}     

\usepackage{enumerate}  

\usepackage{pdfpages}

\usepackage{microtype}

\newcommand*\linkcolours{CornflowerBlue}

\usepackage{bibentry}
\makeatletter
\renewcommand{\bibentry}[1]{\nocite{#1}{\frenchspacing\@nameuse{BR@r@#1\@extra@b@citeb}}}
\makeatother

\usepackage{pgffor}

\usepackage{enumitem}

\newlist{noitemize}{itemize}{1}
\setlist[noitemize]{label={}, labelsep=0pt, leftmargin=0pt}

\usepackage{csquotes}

\usepackage{siunitx}

\usepackage{times}
\usepackage{graphicx}
\usepackage{amssymb}
\usepackage{gensymb}
\usepackage{amsmath}

\usepackage{xurl,hyperref}
\hypersetup{
colorlinks,
linkcolor=\linkcolours,
citecolor=\linkcolours,
filecolor=\linkcolours,
urlcolor=\linkcolours
}

\makeatletter
\newcounter{imagepage}
\newcommand*{\foreachpage}[2]{%
  \begingroup
    \sbox0{\includegraphics{#1}}%
    \xdef\foreachpage@num{\the\pdflastximagepages}%
  \endgroup
  \setcounter{imagepage}{0}%
  \@whilenum\value{imagepage}<\foreachpage@num\do{%
    \stepcounter{imagepage}%
    #2\relax
  }%
}
\makeatother

\usepackage{tocbibind}

\newcommand{\correctionspacing}{\vspace{1mm}}

\phdthesis                         

 \leftchapter                       

\renewcommand{\thesisdepartmentname}{Department of Physics}    

\ifthenelse{\equal{\thesisonlineformat}{true}}{
    \renewcommand{\thesissubmission}{Submitted to the University of Warwick\\
                        for the degree of}
}{
\renewcommand{\thesissubmission}{Submitted to the University of Warwick\\
                        for the degree of}
}

\renewcommand{\thesisauthor}{Jeffrey Mark Ede}  

\renewcommand{\thesismonth}{February}     

\renewcommand{\thesisyear}{2021}      

\renewcommand{\thesistitle}{Advances in Electron Microscopy with Deep Learning}     

\renewcommand{\thesistitletypesize}{\LARGE}   

\renewcommand{\thesisauthorpreviousdegrees}{BSc, MPhys}

\usepackage[toc, abbreviations, acronym, nonumberlist 
    ]{glossaries-extra}
\glssetcategoryattribute{acronym}{nohyperfirst}{true}

\makeatletter
\renewcommand\@pnumwidth{3em}
\makeatother

\makeglossaries
\setabbreviationstyle[acronym]{long-postshort-user}

\newabbreviation{AE}{AE}{Autoencoder}
\newabbreviation{AFM}{AFM}{Atomic Force Microscopy}
\newabbreviation{ALRC}{ALRC}{Adaptive Learning Rate Clipping}
\newabbreviation{ANN}{ANN}{Artificial Neural Network}
\newabbreviation{ASPP}{ASPP}{Atrous Spatial Pyramid Pooling}
\newabbreviation{A-tSNE}{A-tSNE}{Approximate t-Distributed Stochastic Neighbour Embedding}
\newabbreviation{AutoML}{AutoML}{Automatic Machine Learning}
\newabbreviation{Bagged}{Bagged}{Bootstrap Aggregated}
\newabbreviation{bfloat16}{bfloat16}{16 Bit Brain Floating Point}
\newabbreviation{BM3D}{BM3D}{Block-Matching and 3D Filtering}
\newabbreviation{BPTT}{BPTT}{Backpropagation Through Time}
\newabbreviation{CAE}{CAE}{Contractive Autoencoder}
\newabbreviation{CBED}{CBED}{Convergent Beam Electron Diffraction}
\newabbreviation{CBOW}{CBOW}{Continuous Bag-of-Words}
\newabbreviation{CCD}{CCD}{Charge-Coupled Device}
\newabbreviation{cf.}{cf.}{\textit{Confer}}
\newabbreviation{Ch.}{Ch.}{Chapter}
\newabbreviation{CIF}{CIF}{Crystallography Information File}
\newabbreviation{CLRC}{CLRC}{Constant Learning Rate Clipping}
\newabbreviation{CNN}{CNN}{Convolutional Neural Network}
\newabbreviation{COD}{COD}{Crystallography Open Database}
\newabbreviation{COVID-19}{COVID-19}{Coronavirus Disease 2019}
\newabbreviation{CPU}{CPU}{Central Processing Unit}
\newabbreviation{CReLU}{CReLU}{Concatenated Rectified Linear Unit}
\newabbreviation{CTEM}{CTEM}{Conventional Transmission Electron Microscopy}
\newabbreviation{CTF}{CTF}{Contrast Transfer Function}
\newabbreviation{CTRNN}{CTRNN}{Continuous Time Recurrent Neural Network}
\newabbreviation{CUDA}{CUDA}{Compute Unified Device Architecture}
\newabbreviation{cuDNN}{cuDNN}{Compute Unified Device Architecture Deep Neural Network}
\newabbreviation{DAE}{DAE}{Denoising Autoencoder}
\newabbreviation{DALRC}{DALRC}{Doubly Adaptive Learning Rate Clipping}
\newabbreviation{DDPG}{DDPG}{Deep Deterministic Policy Gradients}
\newabbreviation{D-LACBED}{D-LACBED}{Digital Large Angle Convergent Beam Electron Diffraction}
\newabbreviation{DLF}{DLF}{Deep Learning Framework}
\newabbreviation{DLSS}{DLSS}{Deep Learning Supersampling}
\newabbreviation{DNN}{DNN}{Deep Neural Network}
\newabbreviation{DQE}{DQE}{Detective Quantum Efficiency}
\newabbreviation{DSM}{DSM}{Doctoral Skills Module}
\newabbreviation{EBSD}{EBSD}{Electron Backscatter Diffraction}
\newabbreviation{EDX}{EDX}{Energy Dispersive X-Ray}
\newabbreviation{EE}{EE}{Early Exaggeration}
\newabbreviation{EELS}{EELS}{Electron Energy Loss Spectroscopy}
\newabbreviation{e.g.}{e.g.}{\textit{Exempli Gratia}}
\newabbreviation{ELM}{ELM}{Extreme Learning Machine}
\newabbreviation{ELU}{ELU}{Exponential Linear Unit}
\newabbreviation{EM}{EM}{Electron Microscopy}
\newabbreviation{EMDataBank}{EMDataBank}{Electron Microscopy Data Bank}
\newabbreviation{EMPIAR}{EMPIAR}{Electron Microscopy Public Image Archive}
\newabbreviation{EPSRC}{EPSRC}{Engineering and Physical Sciences Research Council}
\newabbreviation{Eqn.}{Eqn.}{Equation}
\newabbreviation{ESN}{ESN}{Echo-State Network}
\newabbreviation{ETDB-Caltech}{ETDB-Caltech}{Caltech Electron Tomography Database}
\newabbreviation{EWR}{EWR}{Exit Wavefunction Reconstruction}
\newabbreviation{FIB-SEM}{FIB-SEM}{Focused Ion Beam Scanning Electron Microscopy}
\newabbreviation{Fig.}{Fig.}{Figure}
\newabbreviation{FFT}{FFT}{Fast Fourier Transform}
\newabbreviation{FNN}{FNN}{Feedforward Neural Network}
\newabbreviation{FPGA}{FPGA}{Field Programmable Gate Array}
\newabbreviation{FT}{FT}{Fourier Transform}
\newabbreviation{FT$^{-1}$}{FT$^{-1}$}{Inverse Fourier Transform}
\newabbreviation{FTIR}{FTIR}{Fourier Transformed Infrared}
\newabbreviation{FTSR}{FTSR}{Focal and Tilt Series Reconstruction}
\newabbreviation{GAN}{GAN}{Generative Adversarial Network}
\newabbreviation{GMS}{GMS}{Gatan Microscopy Suite}
\newabbreviation{GPU}{GPU}{Graphical Processing Unit}
\newabbreviation{GRU}{GRU}{Gated Recurrent Unit}
\newabbreviation{GUI}{GUI}{Graphical User Interface}
\newabbreviation{HPC}{HPC}{High Performance Computing}
\newabbreviation{ICSD}{ICSD}{Inorganic Crystal Structure Database}
\newabbreviation{i.e.}{i.e.}{\textit{Id Est}}
\newabbreviation{i.i.d.}{i.i.d.}{Independent and Identically Distributed}
\newabbreviation{IndRNN}{IndRNN}{Independently Recurrent Neural Network}
\newabbreviation{JSON}{JSON}{Javascript Object Notation}
\newabbreviation{KDE}{KDE}{Kernel Density Estimated}
\newabbreviation{KL}{KL}{Kullback-Leibler}
\newabbreviation{LR}{LR}{Learning Rate}
\newabbreviation{LSTM}{LSTM}{Long Short-Term Memory}
\newabbreviation{LSUV}{LSUV}{Layer-Sequential Unit-Variance}
\newabbreviation{MAE}{MAE}{Mean Absolute Error}
\newabbreviation{MDP}{MDP}{Markov Decision Process}
\newabbreviation{MGU}{MGU}{Minimal Gated Unit}
\newabbreviation{MLP}{MLP}{Multilayer Perceptron}
\newabbreviation{MPAGS}{MPAGS}{Midlands Physics Alliance Graduate School}
\newabbreviation{MTRNN}{MTRNN}{Multiple Timescale Recurrent Neural Network}
\newabbreviation{MSE}{MSE}{Mean Squared Error}
\newabbreviation{N.B.}{N.B.}{\textit{Nota Bene}}
\newabbreviation{NiN}{NiN}{Network-in-Network}
\newabbreviation{NIST}{NIST}{National Institute of Standards and Technology}
\newabbreviation{NMR}{NMR}{Nuclear Magnetic Resonance}
\newabbreviation{NNEF}{MSE}{Neural Network Exchange Format}
\newabbreviation{NTM}{NTM}{Neural Turing Machine}
\newabbreviation{ONNX}{ONNX}{Open Neural Network Exchange}
\newabbreviation{OpenCL}{OpenCL}{Open Computing Language}
\newabbreviation{OU}{OU}{Ornstein-Uhlenbeck}
\newabbreviation{PCA}{PCA}{Principal Component Analysis}
\newabbreviation{PDF}{PDF}{Probability Density Function \textbf{or} Portable Document Format}
\newabbreviation{PhD}{PhD}{Doctor of Philosophy}
\newabbreviation{POMDP}{POMDP}{Partially Observed Markov Decision Process}
\newabbreviation{PReLU}{PReLU}{Parametric Rectified Linear Unit}
\newabbreviation{PSO}{PSO}{Particle Swarm Optimization}
\newabbreviation{RADAM}{RADAM}{Rectified ADAM}
\newabbreviation{RAM}{RAM}{Random Access Memory}
\newabbreviation{RBF}{RBF}{Radial Basis Function}
\newabbreviation{RDPG}{RDPG}{Recurrent Deterministic Policy Gradients}
\newabbreviation{ReLU}{ReLU}{Rectified Linear Unit}
\newabbreviation{REM}{REM}{Reflection Electron Microscopy}
\newabbreviation{RHEED}{RHEED}{Reflection high-energy electron diffraction}
\newabbreviation{RHEELS}{RHEELS}{Reflection High Electron Energy Loss Spectroscopy}
\newabbreviation{RL}{RL}{Reinforcement Learning}
\newabbreviation{RMLP}{RMLP}{Recurrent Multilayer Perceptron}
\newabbreviation{RMS}{RMS}{Root Mean Squared}
\newabbreviation{RNN}{RNN}{Recurrent Neural Network}
\newabbreviation{RReLU}{RReLU}{Randomized Leaky Rectified Linear Unit}
\newabbreviation{RTP}{RTP}{Research Technology Platform}
\newabbreviation{RWI}{RWI}{Random Walk Initialization}
\newabbreviation{SAE}{SAE}{Sparse Autoencoder}
\newabbreviation{SELU}{SELU}{Scaled Exponential Linear Unit}
\newabbreviation{SEM}{SEM}{Scanning Electron Microscopy}
\newabbreviation{SGD}{SGD}{Stochastic Gradient Descent}
\newabbreviation{SNE}{SNE}{Stochastic Neighbour Embedding}
\newabbreviation{SNN}{SNN}{Self-Normalizing Neural Network}
\newabbreviation{SPLEEM}{SPLEEM}{Spin-Polarized Low-Energy Electron Microscopy}
\newabbreviation{SSIM}{SSIM}{Structural Similarity Index Measure}
\newabbreviation{STM}{STM}{Scanning Tunnelling Microscopy}
\newabbreviation{SVD}{SVD}{Singular Value Decomposition}
\newabbreviation{SVM}{SVM}{Support Vector Machine}
\newabbreviation{TEM}{TEM}{Transmission Electron Microscopy}
\newabbreviation{TIFF}{TIFF}{Tag Image File Format}
\newabbreviation{TPU}{TPU}{Tensor Processing Unit}
\newabbreviation{tSNE}{tSNE}{t-Distributed Stochastic Neighbour Embedding}
\newabbreviation{TV}{TV}{Total Variation}
\newabbreviation{URL}{URL}{Uniform Resource Locator}
\newabbreviation{US-tSNE}{US-tSNE}{Uniformly Separated t-Distributed Stochastic Neighbour Embedding}
\newabbreviation{VAE}{VAE}{Variational Autoencoder}
\newabbreviation{VAE-GAN}{VAE-GAN}{Variational Autoencoder Generative Adversarial Network}
\newabbreviation{VBN}{VBN}{Virtual Batch Normalization}
\newabbreviation{VGG}{VGG}{Visual Geometry Group}
\newabbreviation{WDS}{WDS}{Wavelength Dispersive Spectroscopy}
\newabbreviation{WEKA}{WEKA}{Waikato Environment for Knowledge Analysis}
\newabbreviation{WEMD}{WEMD}{Warwick Electron Microscopy Datasets}
\newabbreviation{WLEMD}{WLEMD}{Warwick Large Electron Microscopy Datasets}
\newabbreviation{w.r.t.}{w.r.t.}{With Respect To}
\newabbreviation{XAI}{XAI}{Explainable Artificial Intelligence}
\newabbreviation{XPS}{XPS}{X-Ray Photoelectron Spectroscopy}
\newabbreviation{XRD}{XRD}{X-Ray Diffraction}
\newabbreviation{XRF}{XRF}{X-Ray Fluorescence}

\begin{document}

\nobibliography*

\thesistitlecolourpage           

\pagenumbering{roman} 

\tableofcontents                     

\label{listofabbreviations}
\printunsrtglossary[type=abbreviations,title=List of Abbreviations,toctitle=List of Abbreviations,style=long]

              
\let\oldcite\cite
\newcommand*{\DoNothing}[1]{}%
\renewcommand{\cite}{\DoNothing}

\begin{thesislistoffigures}

\noindent Preface

\begin{itemize}
    \item[1.] Connections between publications covered by chapters of this thesis. An arrow from chapter~$x$ to chapter~$y$ indicates that results covered by chapter~$y$ depend on results covered by chapter~$x$. Labels indicate types of research outputs associated with each chapter, and total connections to and from chapters.
\end{itemize}

\noindent Chapter 1 \quad Review: Deep Learning in Electron Microscopy

\begin{itemize}
    \item[1.] Example applications of a noise-removal DNN to instances of Poisson noise applied to 512$\times$512 crops from TEM images. Enlarged 64$\times$64 regions from the top left of each crop are shown to ease comparison. This figure is adapted from our earlier work\cite{ede2018improvingarxiv} under a Creative Commons Attribution 4.0\cite{cc2020by} license.
    \item[2.] Example applications of DNNs to restore 512$\times$512 STEM images from sparse signals. Training as part of a generative adversarial network\cite{gui2020review, saxena2020generative, pan2019recent, wang2019generative} yields more realistic outputs than training a single DNN with mean squared errors. Enlarged 64$\times$64 regions from the top left of each crop are shown to ease comparison. a) Input is a Gaussian blurred 1/20 coverage spiral\cite{ede2020partial}. b) Input is a 1/25 coverage grid\cite{ede2019deep}. This figure is adapted from our earlier works under Creative Commons Attribution 4.0\cite{cc2020by} licenses.
    \item[3.] Example applications of a semantic segmentation DNN to STEM images of steel to classify dislocation locations. Yellow arrows mark uncommon dislocation lines with weak contrast, and red arrows indicate that fixed widths used for dislocation lines are sometimes too narrow to cover defects. This figure is adapted with permission\cite{roberts2019deep} under a Creative Commons Attribution 4.0\cite{cc2020by} license.
    \item[4.] Example applications of a DNN to reconstruct phases of exit wavefunction from intensities of single TEM images. Phases in $[-\pi, \pi)$ rad are depicted on a linear greyscale from black to white, and Miller indices label projection directions. This figure is adapted from our earlier work\cite{ede2020exit} under a Creative Commons Attribution 4.0\cite{cc2020by} license.
    \item[5.] Reciprocity of TEM and STEM electron optics.
    \item[6.] Numbers of results per year returned by Dimensions.ai abstract searches for SEM, TEM, STEM, STM and REM qualitate their popularities. The number of results for 2020 is extrapolated using the mean rate before 14th July 2020.
    \item[7.] Visual comparison of various normalization methods highlighting regions that they normalize. Regions can be normalized across batch, feature and other dimensions, such as height and width.
    \item[8.] Visualization of convolutional layers. a) Traditional convolutional layer where output channels are sums of biases and convolutions of weights with input channels. b) Depthwise separable convolutional layer where depthwise convolutions compute one convolution with weights for each input channel. Output channels are sums of biases and pointwise convolutions weights with depthwise channels.
    \item[9.] Two 96$\times$96 electron micrographs a) unchanged, and filtered by b) a 5$\times$5 symmetric Gaussian kernel with a 2.5 px standard deviation, c) a 3$\times$3 horizontal Sobel kernel, and d) a 3$\times$3 vertical Sobel kernel. Intensities in a) and b) are in [0, 1], whereas intensities in c) and d) are in [-1, 1].
    \item[10.] Residual blocks where a) one, b) two, and c) three convolutional layers are skipped. Typically, convolutional layers are followed by batch normalization then activation.
    \item[11.] Actor-critic architecture. An actor outputs actions based on input states. A critic then evaluates action-state pairs to predict losses.
    \item[12.] Generative adversarial network architecture. A generator learns to produce outputs that look realistic to a discriminator, which learns to predict whether examples are real or generated.
    \item[13.] Architectures of recurrent neural networks with a) long short-term memory (LSTM) cells, and b) gated recurrent units (GRUs).
    \item[14.] Architectures of autoencoders where an encoder maps an input to a latent space and a decoder learns to reconstruct the input from the latent space. a) An autoencoder encodes an input in a deterministic latent space, whereas a b) traditional variational autoencoder encodes an input as means, $\mu$, and standard deviations, $\sigma$, of Gaussian multivariates, $\mu + \sigma \cdot \epsilon$, where $\epsilon$ is a standard normal multivariate.
    \item[15.] Gradient descent. a) Arrows depict steps across one dimension of a loss landscape as a model is optimized by gradient descent. In this example, the optimizer traverses a small local minimum; however, it then gets trapped in a larger sub-optimal local minimum, rather than reaching the global minimum. b) Experimental DNN loss surface for two random directions in parameter space showing many local minima\cite{li2018visualizing}. The image in part b) is reproduced with permission under an MIT license\cite{li2017mitlicense}.
    \item[16.] Inputs that maximally activate channels in GoogLeNet\cite{szegedy2015going} after training on ImageNet\cite{krizhevsky2012imagenet}. Neurons in layers near the start have small receptive fields and discern local features. Middle layers discern semantics recognisable by humans, such as dogs and wheels. Finally, layers at the end of the DNN, near its logits, discern combinations of semantics that are useful for labelling. This figure is adapted with permission\cite{olah2017feature} under a Creative Commons Attribution 4.0\cite{cc2020by} license.
\end{itemize}

\noindent Chapter 2 \quad  Warwick Electron Microscopy Datasets

\begin{itemize}
    \item[1.] Simplified VAE architecture. a) An encoder outputs means, $\boldsymbol\mu$, and standard deviations, $\boldsymbol\sigma$, to parameterize multivariate normal distributions, $\textbf{z} \sim \text{N}(\boldsymbol\mu, \boldsymbol\sigma)$. b) A generator predicts input images from $\textbf{z}$.
    \item[2.] Images at 500 randomly selected points in two-dimensional tSNE visualizations of 19769 96$\times$96 crops from STEM images for various embedding methods. Clustering is best in a) and gets worse in order a)$\rightarrow$b)$\rightarrow$c)$\rightarrow$d).
    \item[3.] Two-dimensional tSNE visualization of 64-dimensional VAE latent spaces for 19769 STEM images that have been downsampled to 96$\times$96. The same grid is used to show a) map points and b) images at 500 randomly selected points.
    \item[4.] Two-dimensional tSNE visualization of 64-dimensional VAE latent spaces for 17266 TEM images that have been downsampled to 96$\times$96. The same grid is used to show a) map points and b) images at 500 randomly selected points.
\end{itemize}

\noindent Chapter 2 \quad  Supplementary Information: Warwick Electron Microscopy Datasets

\begin{itemize}
    \item[S1.] Two-dimensional tSNE visualization of the first 50 principal components of 19769 STEM images that have been downsampled to 96$\times$96. The same grid is used to show a) map points and b) images at 500 randomly selected points.
    \item[S2.] Two-dimensional tSNE visualization of the first 50 principal components of 19769 96$\times$96 crops from STEM images. The same grid is used to show a) map points and b) images at 500 randomly selected points.
    \item[S3.] Two-dimensional tSNE visualization of the first 50 principal components of 17266 TEM images that have been downsampled to 96$\times$96. The same grid is used to show a) map points and b) images at 500 randomly selected points.
    \item[S4.] Two-dimensional tSNE visualization of the first 50 principal components of 36324 exit wavefunctions that have been downsampled to 96$\times$96. Wavefunctions were simulated for thousands of materials and a large range of physical hyperparameters. The same grid is used to show a) map points and b) wavefunctions at 500 randomly selected points. Red and blue colour channels show real and imaginary components, respectively.
    \item[S5.] Two-dimensional tSNE visualization of the first 50 principal components of 11870 exit wavefunctions that have been downsampled to 96$\times$96. Wavefunctions were simulated for thousands of materials and a small range of physical hyperparameters. The same grid is used to show a) map points and b) wavefunctions at 500 randomly selected points. Red and blue colour channels show real and imaginary components, respectively.
    \item[S6.] Two-dimensional tSNE visualization of the first 50 principal components of 4825 exit wavefunctions that have been downsampled to 96$\times$96. Wavefunctions were simulated for thousands of materials and a small range of physical hyperparameters. The same grid is used to show a) map points and b) wavefunctions at 500 randomly selected points. Red and blue colour channels show real and imaginary components, respectively.
    \item[S7.] Two-dimensional tSNE visualization of means parameterized by 64-dimensional VAE latent spaces for 19769 STEM images that have been downsampled to 96$\times$96. The same grid is used to show a) map points and b) images at 500 randomly selected points.
    \item[S8.] Two-dimensional tSNE visualization of means parameterized by 64-dimensional VAE latent spaces for 19769 96$\times$96 crops from STEM images. The same grid is used to show a) map points and b) images at 500 randomly selected points.
    \item[S9.] Two-dimensional tSNE visualization of means parameterized by 64-dimensional VAE latent spaces for 19769 TEM images that have been downsampled to 96$\times$96. The same grid is used to show a) map points and b) images at 500 randomly selected points.
    \item[S10.] Two-dimensional tSNE visualization of means and standard deviations parameterized by 64-dimensional VAE latent spaces for 19769 96$\times$96 crops from STEM images. The same grid is used to show a) map points and b) images at 500 randomly selected points.
    \item[S11.] Two-dimensional uniformly separated tSNE visualization of 64-dimensional VAE latent spaces for 19769 96$\times$96 crops from STEM images.
    \item[S12.] Two-dimensional uniformly separated tSNE visualization of 64-dimensional VAE latent spaces for 19769 STEM images that have been downsampled to 96$\times$96.
    \item[S13.] Two-dimensional uniformly separated tSNE visualization of 64-dimensional VAE latent spaces for 17266 TEM images that have been downsampled to 96$\times$96.
    \item[S14.] Examples of top-5 search results for 96$\times$96 TEM images. Euclidean distances between $\boldsymbol\mu$ encoded for search inputs and results are smaller for more similar images.
    \item[S15.] Examples of top-5 search results for 96$\times$96 STEM images. Euclidean distances between $\boldsymbol\mu$ encoded for search inputs and results are smaller for more similar images.
\end{itemize}

\noindent Chapter 3 \quad Adaptive Learning Rate Clipping Stabilizes Learning

\begin{itemize}
    \item[1.] Unclipped learning curves for 2$\times$ CIFAR-10 supersampling with batch sizes 1, 4, 16 and 64 with and without adaptive learning rate clipping of losses to 3 standard deviations above their running means. Training is more stable for squared errors than quartic errors. Learning curves are 500 iteration boxcar averaged.
    \item[2.] Unclipped learning curves for 2$\times$ CIFAR-10 supersampling with ADAM and SGD optimizers at stable and unstably high learning rates, $\eta$. Adaptive learning rate clipping prevents loss spikes and decreases errors at unstably high learning rates. Learning curves are 500 iteration boxcar averaged.
    \item[3.] Neural network completions of 512$\times$512 scanning transmission electron microscopy images from 1/20 coverage blurred spiral scans.
    \item[4.] Outer generator losses show that ALRC and Huberization stabilize learning. ALRC lowers final mean squared error (MSE) and Huberized MSE losses and accelerates convergence. Learning curves are 2500 iteration boxcar averaged.
    \item[5.] Convolutional image 2$\times$ supersampling network with three skip-2 residual blocks.
    \item[6.] Two-stage generator that completes 512$\times$512 micrographs from partial scans. A dashed line indicates that the same image is input to the inner and outer generator. Large scale features developed by the inner generator are locally enhanced by the outer generator and turned into images. An auxiliary inner generator trainer restores images from inner generator features to provide direct feedback.
\end{itemize}

\noindent Chapter 4 \quad Partial Scanning Transmission Electron Microscopy with Deep Learning

\begin{itemize}
    \item[1.] Examples of Archimedes spiral (top) and jittered gridlike (bottom) 512$\times$512 partial scan paths for 1/10, 1/20, 1/40, and 1/100 px coverage.
    \item[2.] Simplified multiscale generative adversarial network. An inner generator produces large-scale features from inputs. These are mapped to half-size completions by a trainer network and recombined with the input to generate full-size completions by an outer generator. Multiple discriminators assess multiscale crops from input images and full-size completions. This figure was created with Inkscape\cite{inkscape}.
    \item[3.] Adversarial and non-adversarial completions for 512$\times$512 test set 1/20 px coverage blurred spiral scan inputs. Adversarial completions have realistic noise characteristics and structure whereas non-adversarial completions are blurry. The bottom row shows a failure case where detail is too fine for the generator to resolve. Enlarged 64$\times$64 regions from the top left of each image are inset to ease comparison, and the bottom two rows show non-adversarial generators outputting more detailed features nearer scan paths.
    \item[4.] Non-adversarial generator outputs for 512$\times$512 1/20 px coverage blurred spiral and gridlike scan inputs. Images with predictable patterns or structure are accurately completed. Circles accentuate that generators cannot reliably complete unpredictable images where there is no information. This figure was created with Inkscape\cite{inkscape}.
    \item[5.] Generator mean squared errors (MSEs) at each output pixel for 20000 512$\times$512 1/20 px coverage test set images. Systematic errors are lower near spiral paths for variants of MSE training, and are less structured for adversarial training. Means, $\mu$, and standard deviations, $\sigma$, of all pixels in each image are much higher for adversarial outputs. Enlarged 64$\times$64 regions from the top left of each image are inset to ease comparison, and to show that systematic errors for MSE training are higher near output edges.
    \item[6.] Test set root mean squared (RMS) intensity errors for spiral scans in $[0, 1]$ selected with binary masks. a) RMS errors decrease with increasing electron probe coverage, and are higher than deep learning supersampling\cite{ede2019deep} (DLSS) errors. b) Frequency distributions of 20000 test set RMS errors for 100 bins in $[0, 0.224]$ and scan coverages in the legend.
\end{itemize}

\noindent Chapter 4 \quad Supplementary Information: Partial Scanning Transmission Electron Microscopy with Deep Learning

\begin{itemize}
    \item[S1.] Discriminators examine random $w$$\times$$w$ crops to predict whether complete scans are real or generated. Generators are trained by multiple discriminators with different $w$. This figure was created with Inkscape\cite{inkscape}.
    \item[S2.] Two-stage generator that completes 512$\times$512 micrographs from partial scans. A dashed line indicates that the same image is input to the inner and outer generator. Large scale features developed by the inner generator are locally enhanced by the outer generator and turned into images. An auxiliary trainer network restores images from inner generator features to provide direct feedback. This figure was created with Inkscape\cite{inkscape}.
    \item[S3.] Learning curves. a) Training with an auxiliary inner generator trainer stabilizes training, and converges to lower than two-stage training with fine tuning. b) Concatenating beam path information to inputs decreases losses. Adding symmetric residual connections between strided inner generator convolutions and transpositional convolutions increases losses. c) Increasing sizes of the first inner and outer generator convolutional kernels does not decrease losses. d) Losses are lower after more interations, and a learning rate (LR) of 0.0004; rather than 0.0002. Labels indicate inner generator iterations - outer generator iterations - fine tuning iterations, and k denotes multiplication by 1000 e) Adaptive learning rate clipped quartic validation losses have not diverged from training losses after $10^6$ iterations. f) Losses are lower for outputs in [0, 1] than for outputs in [-1, 1] if leaky ReLU activation is applied to generator outputs.
    \item[S4.] Learning curves. a) Making all convolutional kernels 3$\times$3, and not applying leaky ReLU activation to generator outputs does not increase losses. b) Nearest neighbour infilling decreases losses. Noise was not added to low duration path segments for this experiment. c) Losses are similar whether or not extra noise is added to low-duration path segments. d) Learning is more stable and converges to lower errors at lower learning rates (LRs). Losses are lower for spirals than grid-like paths, and lowest when no noise is added to low-intensity path segments. e) Adaptive momentum-based optimizers, ADAM and RMSProp, outperform non-adaptive momentum optimizers, including Nesterov-accelerated momentum. ADAM outperforms RMSProp; however, training hyperparameters and learning protocols were tuned for ADAM. Momentum values were 0.9. f) Increasing partial scan pixel coverages listed in the legend decreases losses.
    \item[S5.] Adaptive learning rate clipping stabilizes learning, accelerates convergence and results in lower errors than Huberisation. Weighting pixel errors with their running or final mean errors is ineffective.
    \item[S6.] Non-adversarial 512$\times$512 outputs and blurred true images for 1/17.9 px coverage spiral scans selected with binary masks.
    \item[S7.] Non-adversarial 512$\times$512 outputs and blurred true images for 1/27.3 px coverage spiral scans selected with binary masks.
    \item[S8.] Non-adversarial 512$\times$512 outputs and blurred true images for 1/38.2 px coverage spiral scans selected with binary masks.
    \item[S9.] Non-adversarial 512$\times$512 outputs and blurred true images for 1/50.0 px coverage spiral scans selected with binary masks.
    \item[S10.] Non-adversarial 512$\times$512 outputs and blurred true images for 1/60.5 px coverage spiral scans selected with binary masks.
    \item[S11.] Non-adversarial 512$\times$512 outputs and blurred true images for 1/73.7 px coverage spiral scans selected with binary masks.
    \item[S12.] Non-adversarial 512$\times$512 outputs and blurred true images for 1/87.0 px coverage spiral scans selected with binary masks.
\end{itemize}

\noindent Chapter 5 \quad Adaptive Partial Scanning Transmission Electron Microscopy with Reinforcement Learning

\begin{itemize}
    \item[1.] Simplified scan system. a) An example 8$\times$8 partial scan with $T=5$ straight path segments. Each segment in this example has 3 probing positions separated by $d=2^{1/2}$ px, and their starts are labelled by step numbers, $t$. Partial scans are selected from STEM images by sampling image pixels nearest probing positions, even if a nominal probing position is outside an imaging region. b) An actor RNN uses its previous state, action, and an observed path segment to choose the next action at each step. c) A partial scan constructed from actions and observed path segments is completed by a generator CNN.
    \item[2.] Examples of test set 1/23.04 px coverage partial scans, target outputs and generated partial scan completions for 96$\times$96 crops from STEM images. The top four rows show adaptive scans, and the bottom row shows spiral scans. Input partial scans are noisy, whereas target outputs are blurred.
    \item[3.] Learning curves for a)-b) adaptive scan paths chosen by an LSTM or GRU, and fixed spiral and other fixed paths, c) adaptive paths chosen by an LSTM or DNC, d) a range of replay buffer sizes, e) a range of penalties for trying to sample at probing positions over image edges, and f) with and without normalizing or clipping generator losses used for critic training. All learning curves are 2500 iteration boxcar averaged and results in different plots are not directly comparable due to varying experiment settings. Means and standard deviations of test set errors, \enquote{Test: Mean, Std Dev}, are at the ends of labels in graph legends.
\end{itemize}

\noindent Chapter 5 \quad Supplementary Information: Adaptive Partial Scanning Transmission Electron Microscopy with Reinforcement Learning

\begin{itemize}
    \item[S1.] Actor, critic and generator architecture. a) An actor outputs action vectors whereas a critic predicts losses. Dashed lines are for extra components in a DNC. b) A convolutional generator completes partial scans.
    \item[S2.] Learning curves for a) exponentially decayed and exponentially decayed cyclic learning rate schedules, b) actor training with differentiation w.r.t. live or replayed actions, c) images downsampled or cropped from full images to 96$\times$96 with and without additional Sobel losses, d) mean squared error and maximum regional mean squared error loss functions, e) supervision throughout training, supervision only at the start, and no supervision, and f) projection from 128 to 64 hidden units or no projection. All learning curves are 2500 iteration boxcar averaged, and results in different plots are not directly comparable due to varying experiment settings. Means and standard deviations of test set errors, \enquote{Test: Mean, Std Dev}, are at the ends of graph labels.
    \item[S3.] Learning rate optimization. a) Learning rates are increased from $10^{-6.5}$ to $10^{0.5}$ for ADAM and SGD optimization. At the start, convergence is fast for both optimizers. Learning with SGD becomes unstable at learning rates around 2.2$\times$10$^{-5}$, and numerically unstable near 5.8$\times$10$^{-4}$, whereas ADAM becomes unstable around 2.5$\times$10$^{-2}$. b) Training with ADAM optimization for learning rates listed in the legend. Learning is visibly unstable at learning rates of 2.5$\times$10$^{-2.5}$ and 2.5$\times$10$^{-2}$, and the lowest inset validation loss is for a learning rate of 2.5$\times$10$^{-3.5}$. Learning curves in (b) are 1000 iteration boxcar averaged. Means and standard deviations of test set errors, \enquote{Test: Mean, Std Dev}, are at the ends of graph labels.
    \item[S4.] Test set 1/23.04 px coverage adaptive partial scans, target outputs, and generated partial scan completions for 96$\times$96 crops from STEM images.
    \item[S5.] Test set 1/23.04 px coverage adaptive partial scans, target outputs, and generated partial scan completions for 96$\times$96 crops from STEM images.
    \item[S6.] Test set 1/23.04 px coverage spiral partial scans, target outputs, and generated partial scan completions for 96$\times$96 crops from STEM images.
\end{itemize}

\noindent Chapter 6 \quad Improving Electron Micrograph Signal-to-Noise with an Atrous Convolutional Encoder-Decoder

\begin{itemize}
    \item[1.] Simplified network showing how features produced by an Xception backbone are processed. Complex high-level features flow into an atrous spatial pyramid pooling module that produces rich semantic information. This is combined with simple low-level features in a multi-stage decoder to resolve denoised micrographs.
    \item[2.] Mean squared error (MSE) losses of our neural network during training on low dose ($\ll 300$ counts ppx) and fine-tuning for high doses (200-2500 counts ppx). Learning rates (LRs) and the freezing of batch normalization are annotated. Validation losses were calculated using one validation example after every five training batches.
    \item[3.] Gaussian kernel density estimated (KDE)\cite{turlach1993bandwidth, bashtannyk2001bandwidth} MSE and SSIM probability density functions (PDFs) for the denoising methods in table~1. Only the starts of MSE PDFs are shown. MSE and SSIM performances were divided into 200 equispaced bins in [0.0, 1.2] $\times$ $10^{-3}$ and [0.0, 1.0], respectively, for both low and high doses. KDE bandwidths were found using Scott's Rule\cite{scott2015multivariate}.
    \item[4.] Mean absolute errors of our low and high dose networks' 512$\times$512 outputs for 20000 instances of Poisson noise. Contrast limited adaptive histogram equalization\cite{zuiderveld1994contrast} has been used to massively increase contrast, revealing grid-like error variation. Subplots show the top-left 16$\times$16 pixels' mean absolute errors unadjusted. Variations are small and errors are close to the minimum everywhere, except at the edges where they are higher. Low dose errors are in [0.0169, 0.0320]; high dose errors are in [0.0098, 0.0272].
    \item[5.] Example applications of the noise-removal network to instances of Poisson noise applied to 512$\times$512 crops from high-quality micrographs. Enlarged 64$\times$64 regions from the top left of each crop are shown to ease comparison.
    \item[6.] Architecture of our deep convolutional encoder-decoder for electron micrograph denoising. The entry and middle flows develop high-level features that are sampled at multiple scales by the atrous spatial pyramid pooling module. This produces rich semantic information that is concatenated with low-level entry flow features and resolved into denoised micrographs by the decoder.
\end{itemize}

\noindent Chapter 7 \quad Exit Wavefunction Reconstruction from Single Transmission Electron Micrographs with Deep Learning

\begin{itemize}
    \item[1.] Wavefunction propagation. a) An incident wavefunction is perturbed by a projected potential of a material. b) Fourier transforms (FTs) can describe a wavefunction being focused by an objective lens through an objective aperture to a focal plane.
    \item[2.] Crystal structure of In$_{1.7}$K$_2$Se$_8$Sn$_{2.28}$ projected along Miller zone axis [001]. A square outlines a unit cell.
    \item[3.] A convolutional neural network generates $w$$\times$$w$$\times$2 channelwise concatenations of wavefunction components from their amplitudes. Training MSEs are calculated for phase components, before multiplication by input amplitudes.
    \item[4.] A discriminator predicts whether wavefunction components were generated by a neural network.
    \item[5.] Frequency distributions show 19992 validation set mean absolute errors for neural networks trained to reconstruct wavefunctions simulated for multiple materials, multiple materials with restricted simulation hyperparameters, and In$_{1.7}$K$_2$Se$_8$Sn$_{2.28}$. Networks for In$_{1.7}$K$_2$Se$_8$Sn$_{2.28}$ were trained to predict phase components directly; minimising squared errors, and as part of generative adversarial networks. To demonstrate robustness to simulation physics, some validation set errors are shown for $n=1$ and $n=3$ simulation physics. We used up to three validation sets, which cumulatively quantify the ability of a network to generalize to unseen transforms consisting of flips, rotations and translations; simulation hyperparameters, such as thickness and voltage; and materials. A vertical dashed line indicates an expected error of 0.75 for random phases, and frequencies are distributed across 100 bins.
    \item[6.] Training mean absolute errors are similar with and without adaptive learning rate clipping (ALRC). Learning curves are 2500 iteration boxcar averaged.
    \item[7.] Exit wavefunction reconstruction for unseen  NaCl, B$_3$BeLaO$_7$, PbZr$_{0.45}$Ti$_{0.55}$0$_3$, CdTe, and Si input amplitudes, and corresponding crystal structures. Phases in $[-\pi, \pi)$ rad are depicted on a linear greyscale from black to white, and show that output phases are close to true phases. Wavefunctions are cyclically periodic functions of phase so distances between black and white pixels are small. Si is a failure case where phase information is not accurately recovered. Miller indices label projection directions.
\end{itemize}

\noindent Chapter 7 \quad Supplementary Information: Exit Wavefunction Reconstruction from Single Transmission Electron Micrographs with Deep Learning

\begin{itemize}
    \item[S1.] Input amplitudes, target phases and output phases of 224$\times$224 multiple material training set wavefunctions for unseen flips, rotations and translations, and $n=1$ simulation physics.
    \item[S2.] Input amplitudes, target phases and output phases of 224$\times$224 multiple material validation set wavefunctions for seen materials, unseen simulation hyperparameters, and $n=1$ simulation physics.
    \item[S3.] Input amplitudes, target phases and output phases of 224$\times$224 multiple material validation set wavefunctions for unseen materials, unseen simulation hyperparameters, and $n=1$ simulation physics.
    \item[S4.] Input amplitudes, target phases and output phases of 224$\times$224 multiple material training set wavefunctions for unseen flips, rotations and translations, and $n=3$ simulation physics.
    \item[S5.] Input amplitudes, target phases and output phases of 224$\times$224 multiple material validation set wavefunctions for seen materials, unseen simulation hyperparameters, and $n=3$ simulation physics.
    \item[S6.] Input amplitudes, target phases and output phases of 224$\times$224 multiple material validation set wavefunctions for unseen materials, unseen simulation hyperparameters are unseen, and $n=3$ simulation physics.
    \item[S7.] Input amplitudes, target phases and output phases of 224$\times$224 validation set wavefunctions for restricted simulation hyperparameters, and $n=3$ simulation physics.
    \item[S8.] Input amplitudes, target phases and output phases of 224$\times$224 validation set wavefunctions for restricted simulation hyperparameters, and $n=3$ simulation physics.
    \item[S9.] Input amplitudes, target phases and output phases of 224$\times$224 In$_{1.7}$K$_2$Se$_8$Sn$_{2.28}$ training set wavefunctions for unseen flips, rotations and translations, and $n=1$ simulation physics.
    \item[S10.] Input amplitudes, target phases and output phases of 224$\times$224 In$_{1.7}$K$_2$Se$_8$Sn$_{2.28}$ validation set wavefunctions for unseen simulation hyperparameters, and $n=1$ simulation physics.
    \item[S11.] Input amplitudes, target phases and output phases of 224$\times$224 validation set wavefunctions for unseen simulation hyperparameters and materials, and $n=1$ simulation physics. The generator was trained with In$_{1.7}$K$_2$Se$_8$Sn$_{2.28}$ wavefunctions.
    \item[S12.] Input amplitudes, target phases and output phases of 224$\times$224 In$_{1.7}$K$_2$Se$_8$Sn$_{2.28}$ training set wavefunctions for unseen flips, rotations and translations, and $n=1$ simulation physics.
    \item[S13.] Input amplitudes, target phases and output phases of 224$\times$224 In$_{1.7}$K$_2$Se$_8$Sn$_{2.28}$ validation set wavefunctions for unseen simulation hyperparameters, and $n=3$ simulation physics.
    \item[S14.] Input amplitudes, target phases and output phases of 224$\times$224 validation set wavefunctions for unseen simulation hyperparameters and materials, and $n=3$ simulation physics. The generator was trained with In$_{1.7}$K$_2$Se$_8$Sn$_{2.28}$ wavefunctions.
    \item[S15.] GAN input amplitudes, target phases and output phases of 144$\times$144 In$_{1.7}$K$_2$Se$_8$Sn$_{2.28}$ validation set wavefunctions for unseen flips, rotations and translations, and $n=1$ simulation physics.
    \item[S16.] GAN input amplitudes, target phases and output phases of 144$\times$144 In$_{1.7}$K$_2$Se$_8$Sn$_{2.28}$ validation set wavefunctions for unseen simulation hyperparameters, and $n=1$ simulation physics.
    \item[S17.] GAN input amplitudes, target phases and output phases of 144$\times$144 In$_{1.7}$K$_2$Se$_8$Sn$_{2.28}$ validation set wavefunctions for unseen flips, rotations and translations, and $n=3$ simulation physics.
    \item[S18.] GAN input amplitudes, target phases and output phases of 144$\times$144 In$_{1.7}$K$_2$Se$_8$Sn$_{2.28}$ validation set wavefunctions for unseen simulation hyperparameters, and $n=3$ simulation physics.
\end{itemize}

\end{thesislistoffigures}

\begin{thesislistoftables}

\noindent Preface

\begin{itemize}
    \item[1.] Word counts for papers included in thesis chapters, the remainder of the thesis, and the complete thesis.
\end{itemize}

\noindent Chapter 1 \quad Review: Deep Learning in Electron Microscopy

\begin{itemize}
    \item[1.] Deep learning frameworks with programming interfaces. Most frameworks have open source code and many support multiple programming languages.
    \item[2.] Microjob service platforms. The size of typical tasks varies for different platforms and some platforms specialize in preparing machine learning datasets.
\end{itemize}

\noindent Chapter 2 \quad  Warwick Electron Microscopy Datasets

\begin{itemize}
    \item[1.] Examples and descriptions of STEM images in our datasets. References put some images into context to make them more tangible to unfamiliar readers.
    \item[2.] Examples and descriptions of TEM images in our datasets. References put some images into context to make them more tangible to unfamiliar readers.
\end{itemize}

\noindent Chapter 2 \quad  Supplementary Information: Warwick Electron Microscopy Datasets

\begin{itemize}
    \item[S1.] To ease comparison, we have tabulated figure numbers for tSNE visualizations. Visualizations are for principal components, VAE latent space means, and VAE latent space means weighted by standard deviations.
\end{itemize}

\noindent Chapter 3 \quad Adaptive Learning Rate Clipping Stabilizes Learning

\begin{itemize}
    \item[1.] Adaptive learning rate clipping (ALRC) for losses 2, 3, 4 and $\infty$ running standard deviations above their running means for batch sizes 1, 4, 16 and 64. ARLC was not applied for clipping at $\infty$. Each squared and quartic error mean and standard deviation is for the means of the final 5000 training errors of 10 experiments. ALRC lowers errors for unstable quartic error training at low batch sizes and otherwise has little effect. Means and standard deviations are multiplied by 100.
    \item[2.] Means and standard deviations of 20000 unclipped test set MSEs for STEM supersampling networks trained with various learning rate clipping algorithms and clipping hyperparameters, $n^{\uparrow}$ and $n^{\downarrow}$, above and below, respectively.
\end{itemize}

\noindent Chapter 4 \quad Partial Scanning Transmission Electron Microscopy with Deep Learning

\begin{itemize}
    \item[1.] Means and standard deviations of pixels in images created by takings means of 20000 512$\times$512 test set squared difference images with intensities in [-1, 1] for methods to decrease systematic spatial error variation. Variances of Laplacians were calculated after linearly transforming mean images to unit variance.
\end{itemize}

\noindent Chapter 6 \quad Improving Electron Micrograph Signal-to-Noise with an Atrous Convolutional Encoder-Decoder

\begin{itemize}
    \item[1.] Mean MSE and SSIM for several denoising methods applied to 20000 instances of Poisson noise and their standard errors. All methods were implemented with default parameters. Gaussian: 3$\times$3 kernel with a 0.8 px standard deviation. Bilateral: 9$\times$9 kernel with radiometric and spatial scales of 75 (scales below 10 have little effect while scales above 150 cartoonize images). Median: 3$\times$3 kernel. Wiener: no parameters. Wavelet: BayesShrink adaptive wavelet soft-thresholding with wavelet detail coefficient thresholds estimated using \cite{donoho1994ideal}. Chambolle and Bregman TV: iterative total-variation (TV) based denoising\cite{chambolle2004algorithm, goldstein2009split, getreuer2012rudin}, both with denoising weights of 0.1 and applied until the fractional change in their cost function fell below $2.0\times 10^{-4}$ or they reached 200 iterations. Times are for 1000 examples on a 3.4 GHz i7-6700 processor and 1 GTX 1080 Ti GPU, except for our neural network time, which is for 20000 examples.
\end{itemize}

\noindent Chapter 7 \quad Exit Wavefunction Reconstruction from Single Transmission Electron Micrographs with Deep Learning

\begin{itemize}
    \item[1.] New datasets containing 98340 wavefunctions simulated with clTEM are split into training, unseen, validation, and test sets. Unseen wavefunctions are simulated for training set materials with different simulation hyperparameters. Kirkland potential summations were calculated with $n=3$ or truncated to $n=1$ terms, and dashes (-) indicate subsets that have not been simulated. Datasets have been made publicly available at \cite{warwickem!,  ede2020warwick}.
    \item[2.] Means and standard deviations of 19992 validation set errors for unseen transforms (trans.), simulations hyperparameters (param.) and materials (mater.). All networks outperform a baseline uniform random phase generator for both $n=1$ and $n=3$ simulation physics. Dashes (-) indicate that validation set wavefunctions have not been simulated.
\end{itemize}

\end{thesislistoftables}

\renewcommand{\cite}{\oldcite}

\begin{thesisacknowledgments}        


Most modern research builds on a high variety of intellectual contributions, many of which are often overlooked as there are too many to list. Examples include search engines, programming languages, machine learning frameworks, programming libraries, software development tools, computational hardware, operating systems, computing forums, research archives, and scholarly papers. To help developers with limited familiarity, useful resources for deep learning in electron microscopy are discussed in a review paper covered by ch.~\ref{ch:review} of my thesis. For brevity, these acknowledgments will focus on personal contributions to my development as a researcher.

\begin{itemize}
    \item Thanks go to Jeremy Sloan and Richard Beanland for supervision, internal peer review, and co-authorship. 
    \item Thanks go to my Feedback Supervisors, Emma MacPherson and Jon Duffy, for comments needed to partially fulfil requirements of Doctoral Skills Modules (DSMs).
    \item I am grateful to Marin Alexe and Dong Jik Kim for supervising me during a summer project where I programmed various components of atomic force microscopes. It was when I first realized that I want to be a programmer. Before then, I only thought of programming as something that I did in my spare time. 
    \item I am grateful to James Lloyd-Hughes for supervising me during a summer project where I automated Fourier analysis of ultrafast optical spectroscopy signals. 
    \item I am grateful to my family for their love and support.
\end{itemize}

\noindent As a special note, I first taught myself machine learning by working through Mathematica documentation, implementing every machine learning example that I could find. The practice made use of spare time during a two-week course at the start of my Doctor of Philosophy (PhD) studentship, which was needed to partially fulfil requirements of the Midlands Physics Alliance Graduate School (MPAGS).

\vspace{\baselineskip}
\noindent My Head of Department is David Leadley. My Director of Graduate Studies was Matthew Turner, then James Lloyd-Hughes after Matthew Turner retired.

\vspace{\baselineskip}
\noindent I acknowledge funding from Engineering and Physical Sciences Research Council (EPSRC) grant EP/N035437/1 and EPSRC Studentship 1917382.

\end{thesisacknowledgments}

\begin{thesisdeclaration}        
                                 
\textbf{This thesis is submitted to the University of Warwick in support of my application for the degree of Doctor of Philosophy. It has been composed by myself and has not been submitted in any previous application for any degree.}

\correctionspacing
\noindent \textbf{Parts of this thesis have been published by the author:}

\ifthenelse{\equal{\thesisonlineformat}{true}}{}{
\noindent This thesis is typeset for physical printing and binding. In addition, the following version of this thesis\cite{ede2020advances} is typeset for online dissemination to improve readability. 

\begin{quote}
    \bibentry{ede2020advances}
\end{quote}
}

\noindent The following publications\cite{preprint+ede2020review, ede2020warwick, ede2020adaptive, ede2020partial, preprint+ede2020adaptive_scans, ede2019improving, preprint+ede2020exit, ede2020redacted} are part of my thesis.

\begin{quote}
    \bibentry{preprint+ede2020review} 
    
    \bibentry{ede2020warwick} 
    
    \bibentry{ede2020adaptive}
    
    \bibentry{ede2020partial}
    
    \bibentry{preprint+ede2020adaptive_scans}
    
    \bibentry{ede2019improving}
    
    \bibentry{preprint+ede2020exit}    
    
    \bibentry{ede2020redacted}
\end{quote}

\noindent The following publications\cite{ede2020warwick_supplementary, ede2020partial_supplementary, ede2020adaptive_scans_supplementary, ede2020wavefunctions_supplementary} are part of my thesis. However, they are appendices. 

\begin{quote}
    \bibentry{ede2020warwick_supplementary} 
    
    \bibentry{ede2020partial_supplementary}
    
    \bibentry{ede2020adaptive_scans_supplementary}
    
    \bibentry{ede2020wavefunctions_supplementary}
\end{quote}

\noindent The following publications\cite{preprint+ede2020warwick, datasets_repo, warwickem, preprint+ede2019adaptive, alrc_repo, preprint+ede2019partial, preprint+ede2019deep, spirals_repo, dlss_repo, adaptive_scans_repo, preprint+ede2019improving, denoiser_repo, one-shot_repo} are not part of my thesis. However, they are auxiliary to publications that are part of my thesis.

\begin{quote}
    \bibentry{preprint+ede2020warwick} 
    
    \bibentry{datasets_repo} 
    
    \bibentry{warwickem} 
    
    \bibentry{preprint+ede2019adaptive} 
    
    \bibentry{alrc_repo} 
    
    \bibentry{preprint+ede2019partial}
    
    \bibentry{preprint+ede2019deep}
    
    \bibentry{spirals_repo}
    
    \bibentry{dlss_repo}
    
    \bibentry{adaptive_scans_repo}
    
    \bibentry{preprint+ede2019improving}
    
    \bibentry{denoiser_repo}
    
    \bibentry{one-shot_repo}
\end{quote}

\noindent The following publications\cite{ede2020progress_report, beanland+atlas_repo, ede2020wordcount, ede2020posters, preprint+ede2018autoencoders, kernels+MLPs+Autoencoders_repo, simple+webserver_repo} are not part of my thesis. However, they are referenced by my thesis, or are referenced by or associated with publications that are part of my thesis.

\begin{quote}
    \bibentry{ede2020progress_report}

    \bibentry{beanland+atlas_repo}
    
    \bibentry{ede2020wordcount}
    
    \bibentry{ede2020posters}
    
    \bibentry{preprint+ede2018autoencoders}
    
    \bibentry{kernels+MLPs+Autoencoders_repo}
    
    \bibentry{simple+webserver_repo}
\end{quote}


\noindent All publications were produced during my period of study for the degree of Doctor of Philosophy in Physics at the University of Warwick.

\correctionspacing
\noindent \textbf{The work presented (including data generated and data analysis) was carried out by the author except in the cases outlined below:}

\correctionspacing
\noindent Chapter 1 \quad Review: Deep Learning in Electron Microscopy

\begin{quote}
Jeremy Sloan and Martin Lotz internally reviewed my paper after I published it in the arXiv.
\end{quote}

\noindent Chapter 2 \quad  Warwick Electron Microscopy Datasets

\begin{quote}
Richard Beanland internally reviewed my paper before it was published in the arXiv. Further, Jonathan Peters discussed categories used to showcase typical electron micrographs for readers with limited familiarity. At first, our datasets were openly accessible from my Google Cloud Storage. However, Richard Beanland contacted University of Warwick Information Technology Services to arrange for our datasets to also be openly accessible from University of Warwick data servers. Chris Parkin allocated server resources, advised me on data transfer, and handled administrative issues. In addition, datasets are openly accessible from Zenodo and my Google Drive.

Simulated datasets were created with clTEM multislice simulation software developed by a previous EM group PhD student, Mark Dyson, and maintained by a previous EM group postdoctoral researcher, Jonathan Peters. Jonathan Peters advised me on processing data that I had curated from the Crystallography Open Database (COD) so that it could be input into clTEM simulations. Further, Jonathan Peters and I jointly prepared a script to automate multislice simulations. Finally, Jonathan Peters computed a third of our simulations on his graphical processing units (GPUs). 


Experimental datasets were curated from University of Warwick Electron Microscopy (EM) Research Technology Platform (RTP) dataservers, and contain images collected by dozens of scientists working on hundreds of projects. Data was curated and published with permission of the Director of the EM RTP, Richard Beanland. In addition, data curation and publication were reviewed and approved by Research Data Officers, Yvonne Budden and Heather Lawler. I was introduced to the EM dataservers by Richard Beanland and Jonathan Peters, and my read and write access to the EM dataservers was set up by an EM RTP technician, Steve York.
\end{quote}

\noindent Chapter 3 \quad Adaptive Learning Rate Clipping Stabilizes Learning

\begin{quote}
Richard Beanland internally reviewed my paper after it was published in the arXiv. Martin Lotz later recommend the journal that I published it in. In addition, a Scholarly Communications Manager, Julie Robinson, advised me on finding publication venues and open access funding. I also discussed publication venues with editors of Machine Learning, Melissa Fearon and Peter Flach, and my Centre for Scientific Computing Director, David Quigley.
\end{quote}

\noindent Chapter 4 \quad Partial Scanning Transmission Electron Microscopy with Deep Learning

\begin{quote}
Richard Beanland internally reviewed an initial draft of my paper on partial scanning transmission electron microscopy (STEM). After I published our paper in the arXiv, Richard Beanland contributed most of the content in the first two paragraphs in the introduction of the journal paper. In addition, Richard Beanland and I both copyedited our paper. 

Richard Beanland internally reviewed a paper on uniformly spaced scans after I published it in the arXiv. The uniformly spaced scans paper includes some experiments that we later combined into our partial STEM paper. Further, my experiments followed a preliminary investigation into compressed sensing with fixed randomly spaced masks, which Richard Beanland internally reviewed. 
\end{quote}

\noindent Chapter 5 \quad Adaptive Partial Scanning Transmission Electron Microscopy with Reinforcement Learning

\begin{quote}
Jasmine Clayton, Abdul Mohammed, and Jeremy Sloan internally reviewed my paper after I published it in the arXiv.
\end{quote}

\noindent Chapter 6 \quad Improving Electron Micrograph Signal-to-Noise with an Atrous Convolutional Encoder-Decoder

\begin{quote}
After I published my paper in the arXiv, Richard Beanland internally reviewed it and advised that we publish it in a journal. In addition, Richard Beanland and I both copyedited our paper. 
\end{quote}

\noindent Chapter 7 \quad Exit Wavefunction Reconstruction from Single Transmission Electron Micrographs with Deep Learning

\begin{quote}
Jeremy Sloan internally reviewed an initial draft of our paper. Afterwards, Jeremy Sloan contributed all crystal structure diagrams in our paper. The University of Warwick X-Ray Facility Manager, David Walker, suggested materials to showcase with their crystal structures, and a University of Warwick Research Fellow, Jessica Marshall, internally reviewed a figure showing exit wavefunction reconstructions (EWRs) with the crystal structures. 

Richard Beanland contacted a professor at Humboldt University of Berlin, Christoph Koch, to ask for a DigitalMicrograph plugin, which I used to collect experimental focal series. Further, Richard Beanland helped me get started with focal series measurements, and internally reviewed some of my first focal series. In addition, Richard Beanland internally reviewed our paper. 

Jonathan Peters drafted initial text about clTEM multislice simulations for a section of our paper on \enquote{Exit Wavefunction Datasets}. In addition, Jonathan Peters internally reviewed our paper.
\end{quote}

\correctionspacing
\noindent \textbf{This thesis conforms to regulations governing the examination of higher degrees by research:}

\noindent The following regulations\cite{doctoral_college_regulations, university_calendar_regulations} were used during preparation of this thesis.

\begin{quote}
    \bibentry{doctoral_college_regulations}
    
    \bibentry{university_calendar_regulations}
\end{quote}

\noindent The following guidance\cite{department_of_physics_guidance} was helpful during preparation of this thesis.

\begin{quote}
    \bibentry{department_of_physics_guidance}
\end{quote}

\noindent The following thesis template\cite{latex_style_files} was helpful during preparation of this thesis.

\begin{quote}
    \bibentry{latex_style_files}
\end{quote}

\noindent Thesis structure and content was discussed with my previous Director of Graduate Studies, Matthew Turner, and my current Director of Graduate Studies, James Lloyd-Hughes, after Matthew Turner retired. My thesis was also discussed with my both my previous PhD supervisor, Richard Beanland, and my current PhD supervisor, Jeremy Sloan. My formal thesis plan was then reviewed and approved by both Jeremy Sloan and my feedback supervisor, Emma MacPherson. Finally, my complete thesis was internally reviewed by both Jeremy Sloan and Jasmine Clayton. 

Permission is granted by the Chair of the Board of Graduate Studies, Colin Sparrow, for my thesis appendices to exceed length requirements usually set by the University of Warwick. This is in the understanding that my thesis appendices are not usually crucial to the understanding or examination of my thesis.

\end{thesisdeclaration}



\begin{thesisresearchtraining}

This thesis presents a substantial original investigation of deep learning in electron microscopy. The only researcher in my research group or building with machine learning expertise was myself. This meant that I led the design, implementation, evaluation, and publication of experiments covered by my thesis. Where experiments were collaborative, I both proposed and led the collaboration. 

\end{thesisresearchtraining}

\begin{thesisabstract}               

Following decades of exponential increases in computational capability and widespread data availability, deep learning is readily enabling new science and technology. This thesis starts with a review of deep learning in electron microscopy, which offers a practical perspective aimed at developers with limited familiarity. To help electron microscopists get started with started with deep learning, large new electron microscopy datasets are introduced for machine learning. Further, new approaches to variational autoencoding are introduced to embed datasets in low-dimensional latent spaces, which are used as the basis of electron microscopy search engines. Encodings are also used to investigate electron microscopy data visualization by t-distributed stochastic neighbour embedding. Neural networks that process large electron microscopy images may need to be trained with small batch sizes to fit them into computer memory. Consequently, adaptive learning rate clipping is introduced to prevent learning being destabilized by loss spikes associated with small batch sizes.

This thesis presents three applications of deep learning to electron microscopy. Firstly, electron beam exposure can damage some specimens, so generative adversarial networks were developed to complete realistic images from sparse spiral, gridlike, and uniformly spaced scans. Further, recurrent neural networks were trained by reinforcement learning to dynamically adapt sparse scans to specimens. Sparse scans can decrease electron beam exposure and scan time by 10-100$\times$ with minimal information loss. Secondly, a large encoder-decoder was developed to improve transmission electron micrograph signal-to-noise. Thirdly, conditional generative adversarial networks were developed to recover exit wavefunction phases from single images. Phase recovery with deep learning overcomes existing limitations as it is suitable for live applications and does not require microscope modification. To encourage further investigation, scientific publications and their source files, source code, pretrained models, datasets, and other research outputs covered by this thesis are openly accessible.

\end{thesisabstract}



\begin{thesispreface}

This thesis covers a subset of my scientific papers on advances in electron microscopy with deep learning. The papers were prepared while I was a PhD student at the University of Warwick in support of my application for the degree of PhD in Physics. This thesis reflects on my research, unifies covered publications, and discusses future research directions. My papers are available as part of chapters of this thesis, or from their original publication venues with hypertext and other enhancements. This preface covers my initial motivation to investigate deep learning in electron microscopy, structure and content of my thesis, and relationships between included publications. Traditionally, physics PhD theses submitted to the University of Warwick are formatted for physical printing and binding. However, I have also formatted a copy of my thesis for online dissemination to improve readability\cite{ede2020advances}.

\section{Initial Motivation}

When I started my PhD in October 2017, we were unsure if or how machine learning could be applied to electron microscopy. My PhD was funded by EPSRC Studentship 1917382\cite{uk2020epsrc} titled \enquote{Application of Novel Computing and Data Analysis Methods in Electron Microscopy}, which is associated with EPSRC grant EP/N035437/1\cite{epsrc2016epsrc} titled \enquote{ADEPT -- Advanced Devices by ElectroPlaTing}. As part of the grant, our initial plan was for me to spend a couple of days per week using electron microscopes to analyse specimens sent to the University of Warwick from the University of Southampton, and to invest remaining time developing new computational techniques to help with analysis. However, an additional scientist was not needed to analyse specimens, so it was difficult for me to get electron microscopy training. While waiting for training, I was tasked with automating analysis of digital large angle convergent beam electron diffraction\cite{hubert2019structure} (D-LACBED) patterns. However, we did not have a compelling use case for my D-LACBED software\cite{ede2018beanland, ede2020progress_report}. Further, a more senior PhD student at the University of Warwick, Alexander Hubert, was already investigating convergent beam electron diffraction\cite{hubert2019structure, hart2016electron} (CBED).

My first machine learning research began five months after I started my PhD. Without a clear research direction or specimens to study, I decided to develop artificial neural networks (ANNs) to generate artwork. My dubious plan was to create image processing pipelines for the artwork, which I would replace with electron micrographs when I got specimens to study. However, after investigating artwork generation with randomly initialized multilayer perceptrons\cite{ha2015neural, le2019generate}, then by style transfer\cite{gatys2016image, gatys2015neural}, and then by fast style transfer\cite{johnson2016perceptual}, there were still no specimens for me to study. Subsequently, I was inspired by NVIDIA's research on semantic segmentation\cite{wang2018high} to investigate semantic segmentation with DeepLabv3+\cite{chen2018encoder}. However, I decided that it was unrealistic for me to label a large new electron microscopy dataset for semantic segmentation by myself. Fortunately, I had read about using deep neural networks (DNNs) to reduce image compression artefacts\cite{jiang2017end}, so I wondered if a similar approach based on DeepLabv3+ could improve electron micrograph signal-to-noise. Encouragingly, it would not require time-consuming image labelling. Following a successful investigation into improving signal-to-noise, my first scientific paper\cite{ede2019improving} (ch.~\ref{ch:denoiser}) was submitted a few months later, and my experience with deep learning enabled subsequent investigations.

\begin{figure*}[tbh!]
\includegraphics[width=\textwidth]{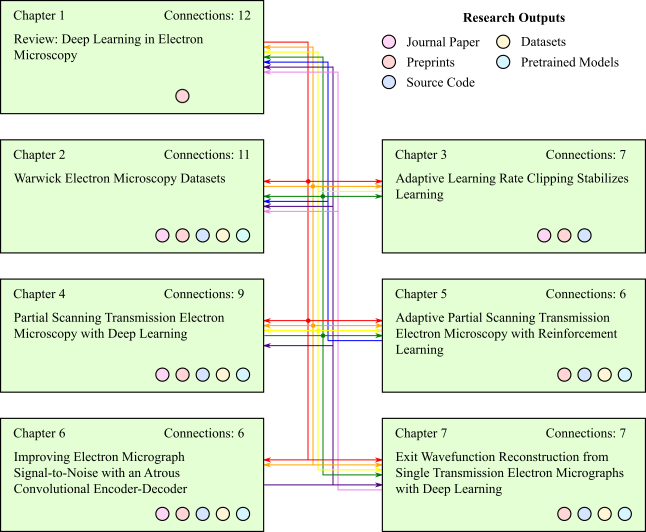}
\caption{ Connections between publications covered by chapters of this thesis. An arrow from chapter~$x$ to chapter~$y$ indicates that results covered by chapter~$y$ depend on results covered by chapter~$x$. Labels indicate types of research outputs associated with each chapter, and total connections to and from chapters. }
\label{fig:publications_relationship}
\end{figure*}

\section{Thesis Structure}

An overview of the first seven chapters in this thesis is presented in fig.~\ref{fig:publications_relationship}. The first chapter is introductory and covers a review of deep learning in electron microscopy, which offers a practical perspective aimed at developers with limited familiarity. The next two chapters are ancillary and cover new datasets and an optimization algorithm used in later chapters. The final four chapters before conclusions cover investigations of deep learning in electron microscopy. Each of the first seven chapter covers a combination of journal papers, preprints, and ancillary outputs such as source code, datasets, and pretrained models, and supplementary information.

At the University of Warwick, physics PhD theses that cover publications\cite{guerin2016connecting, mason2020institutional} are unusual. Instead, most theses are scientific monographs. However, declining impact of monographic theses is long-established\cite{lariviere2008declining}, and I felt that scientific publishing would push me to produce higher-quality research. Moreover, I think that publishing is an essential part of scientific investigation, and external peer reviews\cite{publons20182018, tennant2018state, walker2015emerging, vesper2018peer, tan2019performance} often helped me to improve my papers. Open access to PhD theses increases visibility\cite{ferreras2016providing, kettler2016ways} and enables their use as data mining resources\cite{kettler2016ways, miller2015making}, so digital copies of physics PhD theses are archived by the University of Warwick\cite{university2020theses}. However, archived theses are usually formatted for physical printing and binding. To improve readability, I have also formatted a copy of my thesis for online dissemination\cite{ede2020advances}, which is published in the arXiv\cite{arxiv2020about, ginsparg2011arxiv} with its Latex\cite{pignalberi2019introduction, bransen2018pimp, lamport1994latex} source files.

All my papers were first published as arXiv preprints under Creative Commons Attribution 4.0\cite{cc2020by} licenses, then submitted to journals. As discussed in my review\cite{preprint+ede2020review} (ch.~\ref{ch:review}), advantages of preprint archives\cite{hoy2020rise, fry2019praise, rodriguez2019preprints} include ensuring that research is openly accessible\cite{banks2019answers}, increasing discovery and citations\cite{fraser2020relationship, wang2020impact, furnival2020open, fu2019meta, niyazov2016open}, inviting timely scientific discussion, and raising awareness to reduce unnecessary duplication of research. Empirically, there are no significant textual differences between arXiv preprints and corresponding journal papers\cite{klein2019comparing}. However, journal papers appear to be slightly higher quality than biomedical preprints\cite{carneiro2019comparing, klein2019comparing}, suggesting that formatting and copyediting practices vary between scientific disciplines. Overall, I think that a lack of differences between journal papers and preprints may be a result of publishers separating language editing into premium services\cite{elsevier2020language, iop2020editing, springernature2020author, wiley2020editing}, rather than including extensive language editing in their usual publication processes. Increasing textual quality is correlated with increasing likelihood that an article will be published\cite{roth2019understanding}. However, most authors appear to be performing copyediting themselves to avoid extra fees. 

A secondary benefit of posting arXiv preprints is that their metadata, an article in portable document format\cite{international2017iso, adobe2008iso} (PDF), and any Latex source files are openly accessible. This makes arXiv files easy to reuse, especially if they are published under permissive licenses\cite{arxiv2020license}. For example, open accessibility enabled arXiv files to be curated into a large dataset\cite{clement2019use} that was used to predict future research trends\cite{eger2019predicting}. Further, although there is no requirement for preprints to peer reviewed, preprints can enable early access to papers that have been peer reviewed. As a case in point, all preprints covered by my thesis have been peer reviewed. Further, the arXiv implicitly supports peer review by providing contact details of authors, and I have both given and received feedback about arXiv papers. In addition, open peer review platforms\cite{ross2017open}, such as OpenReview\cite{openreview2020about, soergel2013open}, can be used to explicitly seek peer review. There is also interest in integrating peer review with the arXiv, so a conceptual peer review model has been proposed\cite{wang2019conceptual}. 

\setcounter{table}{0}

\begin{table}[tbh!]
\footnotesize
\centering
\begin{tabular*}{\textwidth}{l@{\extracolsep{\fill}}cccc}
\hline
Description & Words in Text & Words in Figures & Words in Algorithms & Total Words \\
\hline
Review paper in chapter 1 & 15156 & 2680 & 74 & 17910 \\
Ancillary paper in chapter 2 & 4243 & 1360 & 0 & 5603 \\
Ancillary paper in chapter 3 & 2448 & 680 & 344 & 3472 \\
Paper in chapter 4 & 3864 & 1300 & 0 & 5164 \\
Paper in chapter 5 & 3399 & 900 & 440 & 4739 \\
Paper in chapter 6 & 2933 & 1100 & 0 & 4033 \\
Paper in chapter 7 & 4396 & 1240 & 0 & 5636 \\
Remainder of the thesis & 7950 & 280 & 0 & 8230 \\
\hline
\textbf{Complete thesis} & \textbf{44389} & \textbf{9540} & \textbf{858} & \textbf{54787} \\
\hline
\end{tabular*}
\caption{ Word counts for papers included in thesis chapters, the remainder of the thesis, and the complete thesis. }
\label{table:word_count}
\end{table}

This thesis covers a selection of my interconnected scientific papers. Word counts for my papers and covering text are tabulated in table~\ref{table:word_count}. Figures are included in word counts by adding products of nominal word densities and figure areas. However, acknowledgements, references, tables, supplementary information, and similar contents are not included as they do not count towards my thesis length limit of 70000 words. For details, notes on my word counting procedure are openly accessible\cite{ede2020wordcount}. Associated research outputs, such as source code and datasets, are not directly included in my thesis due to format restrictions. Nevertheless, my source code is openly accessible from GitHub\cite{github2020profile}, and archived releases of my source code are openly accessible from Zenodo\cite{zenodo}. In addition, links to openly accessible pretrained models are provided in my source code documentation. Finally, links to openly accessible datasets are in my papers, source code documentation, and datasets paper\cite{ede2020warwick} (ch.~\ref{ch:wemd}).

\section{Connections}

Connections between publications covered by my thesis are shown in fig.~\ref{fig:publications_relationship}. The most connected chapter covers my review paper\cite{preprint+ede2020review} (ch.~\ref{ch:review}). All my papers are connected to my review paper as literature reviews informed their introductions, methodologies, and discussions. My review paper also discusses and builds upon the results of my earlier publications. For example, images published in my earlier papers are reused in my review paper to showcase applications of deep learning in electron microscopy. In addition, my review paper covers Warwick Electron Microscopy Datasets\cite{ede2020warwick} (WEMD, ch.~\ref{ch:wemd}), adaptive learning rate clipping\cite{ede2020adaptive} (ALRC, ch.~\ref{ch:ALRC}), sparse scans for compressed sensing in STEM\cite{ede2020partial} (ch.~\ref{ch:partial_stem}), improving electron microscope signal-to-noise\cite{ede2019improving} (ch.~\ref{ch:denoiser}), and EWR\cite{preprint+ede2020exit} (ch.~\ref{ch:ewr}). Finally, compressed sensing with dynamic scan paths that adapt to specimens\cite{preprint+ede2020adaptive_scans} (ch.~\ref{ch:adaptive_scans}) motivated my review paper sections on recurrent neural networks (RNNs) and reinforcement learning (RL).

The second most connected chapter, ch.~\ref{ch:wemd}, is ancillary and covers WEMD\cite{ede2020warwick}, which include large new datasets of experimental transmission electron microscopy (TEM) images, experimental STEM images, and simulated exit wavefunctions. The TEM images were curated to train an ANN to improve signal-to-noise\cite{ede2019improving} (ch.~\ref{ch:denoiser}) and motivated the proposition of a new approach to EWR\cite{preprint+ede2020exit} (ch.~\ref{ch:ewr}). The STEM images were curated to train ANNs for compressed sensing\cite{ede2020partial} (ch.~\ref{ch:partial_stem}). Training our ANNs with full-size images was impractical with our limited computational resources, so I created dataset variants containing 512$\times$512 crops from full-size images for both the TEM and STEM datasets. However, 512$\times$512 STEM crops were too large to efficiently train RNNs to adapt scan paths\cite{preprint+ede2020adaptive_scans} (ch.~\ref{ch:adaptive_scans}), so I also created 96$\times$96 variants of datasets for rapid initial development. Finally, datasets of exit wavefunctions were simulated as part of our initial investigation into EWR from single TEM images with deep learning\cite{preprint+ede2020exit} (ch.~\ref{ch:ewr}).

The other ancillary chapter, ch.~\ref{ch:ALRC}, covers ALRC\cite{ede2020adaptive}, which was originally published as an appendix in the first version of our partial STEM preprint\cite{preprint+ede2019partial} (ch.~\ref{ch:partial_stem}). The algorithm was developed to stabilize learning of ANNs being developed for partial STEM, which were destabilized by loss spikes when training with a batch size of 1. My aim was to make experiments\cite{ede2020partial_supplementary} easier to compare by preventing learning destabilized by large loss spikes from complicating comparisons. However, ALRC was so effective that I continued to investigate it, increasing the size of the partial STEM appendix. Eventually, the appendix became so large that I decided to turn it into a short paper. To stabilize training with small batch sizes, ALRC was also applied to ANN training for uniformly spaced scans\cite{ede2020partial, preprint+ede2019deep} (ch.~\ref{ch:partial_stem}). In addition, ALRC inspired adaptive loss clipping to stabilize RNN training for adaptive scans\cite{preprint+ede2020adaptive_scans} (ch.~\ref{ch:adaptive_scans}). Finally, I investigated applying ALRC to ANN training for EWR\cite{preprint+ede2020exit} (ch.~\ref{ch:ewr}). However, ALRC did not improve EWR as training with a batch size of 32 was not destabilized by loss spikes.

My experiments with compressed sensing showed that ANN performance varies for different scan paths\cite{ede2020partial} (ch.~\ref{ch:partial_stem}). This motivated the investigation of scan shapes that adapt to specimens as they are scanned\cite{preprint+ede2020adaptive_scans} (ch.~\ref{ch:adaptive_scans}). I had found that ANNs for TEM denoising\cite{ede2019improving} (ch.~\ref{ch:denoiser}) and uniformly spaced sparse scan completion\cite{preprint+ede2019deep} exhibit significant structured systematic error variation, where errors are higher near output edges. Subsequently, I investigated average partial STEM output errors and found that errors increase with increasing distance from scan paths\cite{ede2020partial} (ch.~\ref{ch:partial_stem}). In part, structured systematic error variation in partial STEM\cite{ede2020partial} (ch.~\ref{ch:partial_stem}) motivated my investigation of adaptive scans\cite{preprint+ede2020adaptive_scans} (ch.~\ref{ch:adaptive_scans}) as I reasoned that being able to more closely scan regions where errors would otherwise be highest could decrease mean errors. 

Most of my publications are connected by their source code as it was partially reused in successive experiments. Source code includes scripts to develop ANNs, plot graphs, create images for papers, and typeset with Latex. Following my publication chronology, I partially reused source code created to improve signal-to-noise\cite{ede2019improving} (ch.~\ref{ch:denoiser}) for partial STEM\cite{ede2020partial} (ch.~\ref{ch:partial_stem}). My partial STEM source code was then partially reused for my other investigations. Many of my publications are also connected because datasets curated for my first investigations were reused in my later investigations. For example, improving signal-to-noise\cite{ede2019improving} (ch.~\ref{ch:denoiser}) is connected to EWR\cite{preprint+ede2020exit} (ch.~\ref{ch:ewr}) as the availability of my large dataset of TEM images prompted the proposition of, and may enable, a new approach to EWR. Similarly, partial STEM\cite{ede2020partial} (ch.~\ref{ch:partial_stem}) is connected to adaptive scans\cite{preprint+ede2020adaptive_scans} (ch.~\ref{ch:adaptive_scans}) as my large dataset of STEM images was used to derive smaller datasets used to rapidly develop adaptive scan systems.

\end{thesispreface}

\pagenumbering{arabic} 

\chapter{Review: Deep Learning in Electron Microscopy}\label{ch:review}

\section{Scientific Paper}\label{sec:diss_review}

\noindent This chapter covers the following paper\cite{preprint+ede2020review}.
\begin{quote}
\bibentry{preprint+ede2020review}
\end{quote}

\foreachpage{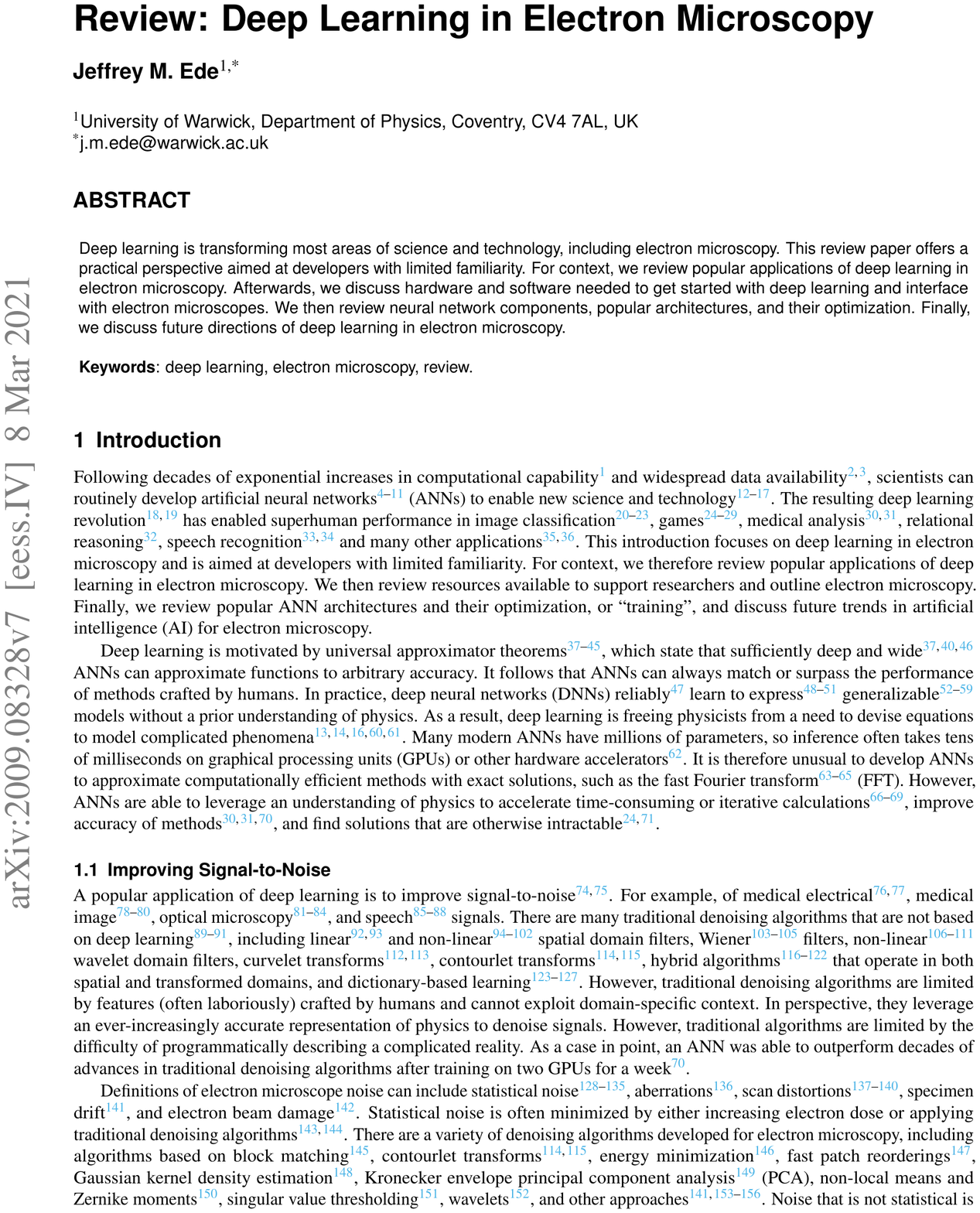}{%
  \ifthenelse{\value{imagepage}>0}{
  \newpage   
  \begingroup 
    \centering
    \includegraphics[
      trim={20mm 22mm 20mm 22.5mm},
      clip,
      page=\value{imagepage},
      width=\textwidth,  
      height=\textheight,
      keepaspectratio,
    ]{arxiv_review}
    \newpage
  \endgroup
  }{}
}

\section{Reflection}

This introductory chapter covers my review paper\cite{lalli2019brief} titled \enquote{Review: Deep Learning in Electron Microscopy}\cite{preprint+ede2020review}. It is the first in-depth review of deep learning in electron microscopy and offers a practical perspective that is aimed at developers with limited familiarity. My review was crafted to be covered by the introductory chapter of my PhD thesis, so focus is placed on my research methodology. Going through its sections in order of appearance, \enquote{Introduction} covers and showcases my earlier research, \enquote{Resources} introduces resources that enabled my research, \enquote{Electron Microscopy} covers how I simulated exit wavefunctions and integrated ANNs with electron microscopes, \enquote{Components} introduces functions used to construct my ANNs, \enquote{Architecture} details ANN archetypes used in my research, \enquote{Optimization} covers how my ANNs were trained, and \enquote{Discussion} offers my perspective on deep learning in electron microscopy.

There are many review papers on deep learning. Some reviews of deep learning  focus on computer science\cite{sengupta2020review, shrestha2019review, alom2019state, lecun2015deep, schmidhuber2015deep}, whereas others focus on specific applications such as computational imaging\cite{barbastathis2019use}, materials science\cite{ge2020deep, wei2019machine, schleder2019dft}, and the physical sciences\cite{carleo2019machine}. As a result, I anticipated that another author might review deep learning in electron microscopy. To avoid my review being easily surpassed, I leveraged my experience to offer practical perspectives and comparative discussions to address common causes of confusion. In addition, content is justified by extensive references to make it easy to use as a starting point for future research. Finally, I was concerned that information about how to get started with deep learning in electron microscopy was fragmented and unclear to unfamiliar developers. This was often problematic when I was asked about getting started with machine learning, and I was especially conscious of it as my friend, Rajesh Patel, asked me for advice when I started writing my review. Consequently, I included a section that introduces useful resources for deep learning in electron microscopy.



\chapter{Warwick Electron Microscopy Datasets}\label{ch:wemd}

\section{Scientific Paper}\label{sec:diss_warwick}

\noindent This paper covers the following paper\cite{ede2020warwick} and its supplementary information\cite{ede2020warwick_supplementary}.
\begin{quote}
\bibentry{ede2020warwick}

\bibentry{ede2020warwick_supplementary}
\end{quote}

\foreachpage{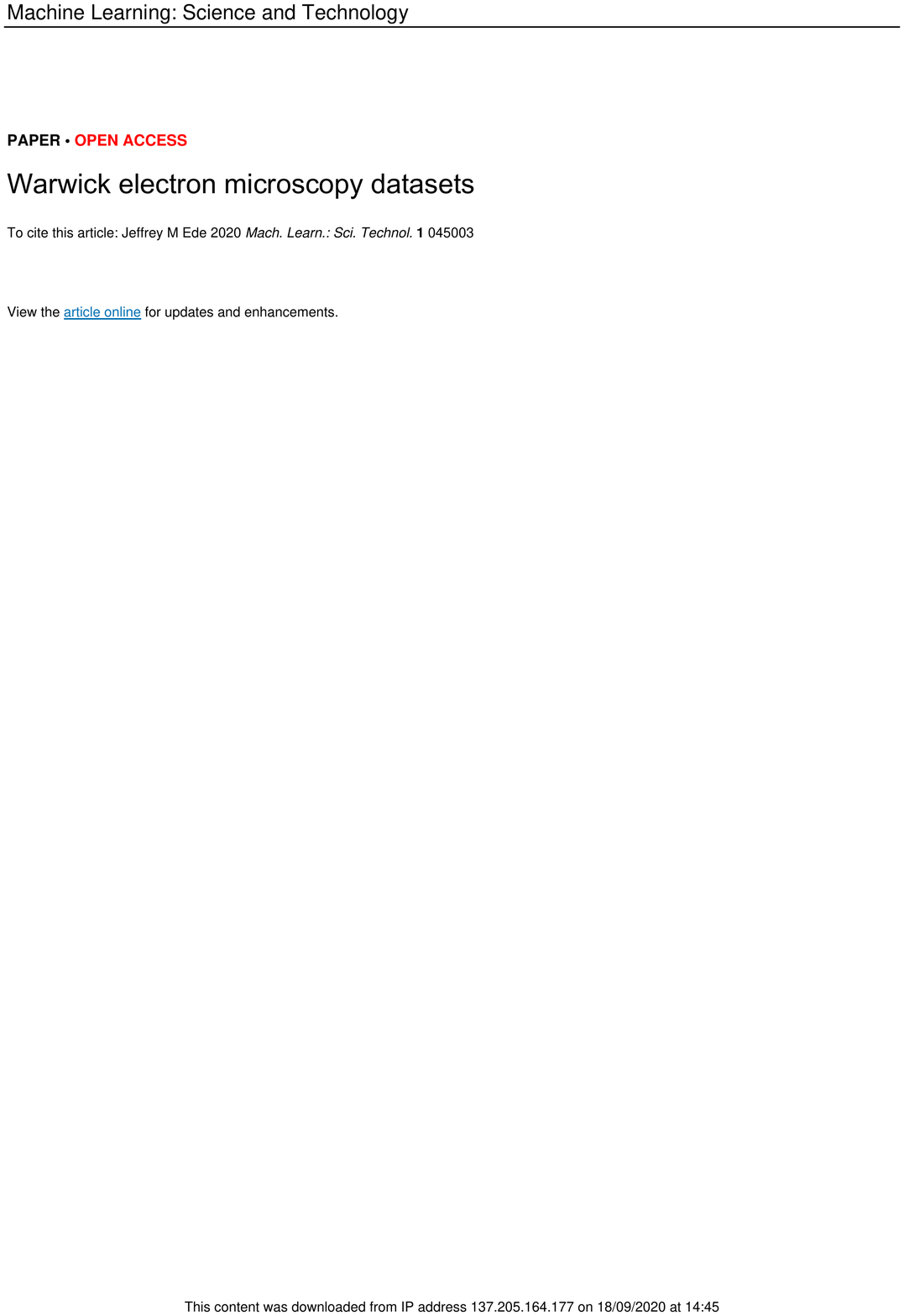}{%
  \ifthenelse{\value{imagepage}>1}{
  \newpage   
  \begingroup 
    \centering
    \includegraphics[
      trim={15.25mm 15.5mm 15.25mm 12.5mm},
      clip,
      page=\value{imagepage},
      width=\textwidth,  
      height=\textheight,
      keepaspectratio,
    ]{paper_wemd}
    \newpage
  \endgroup
  }{}
}

\foreachpage{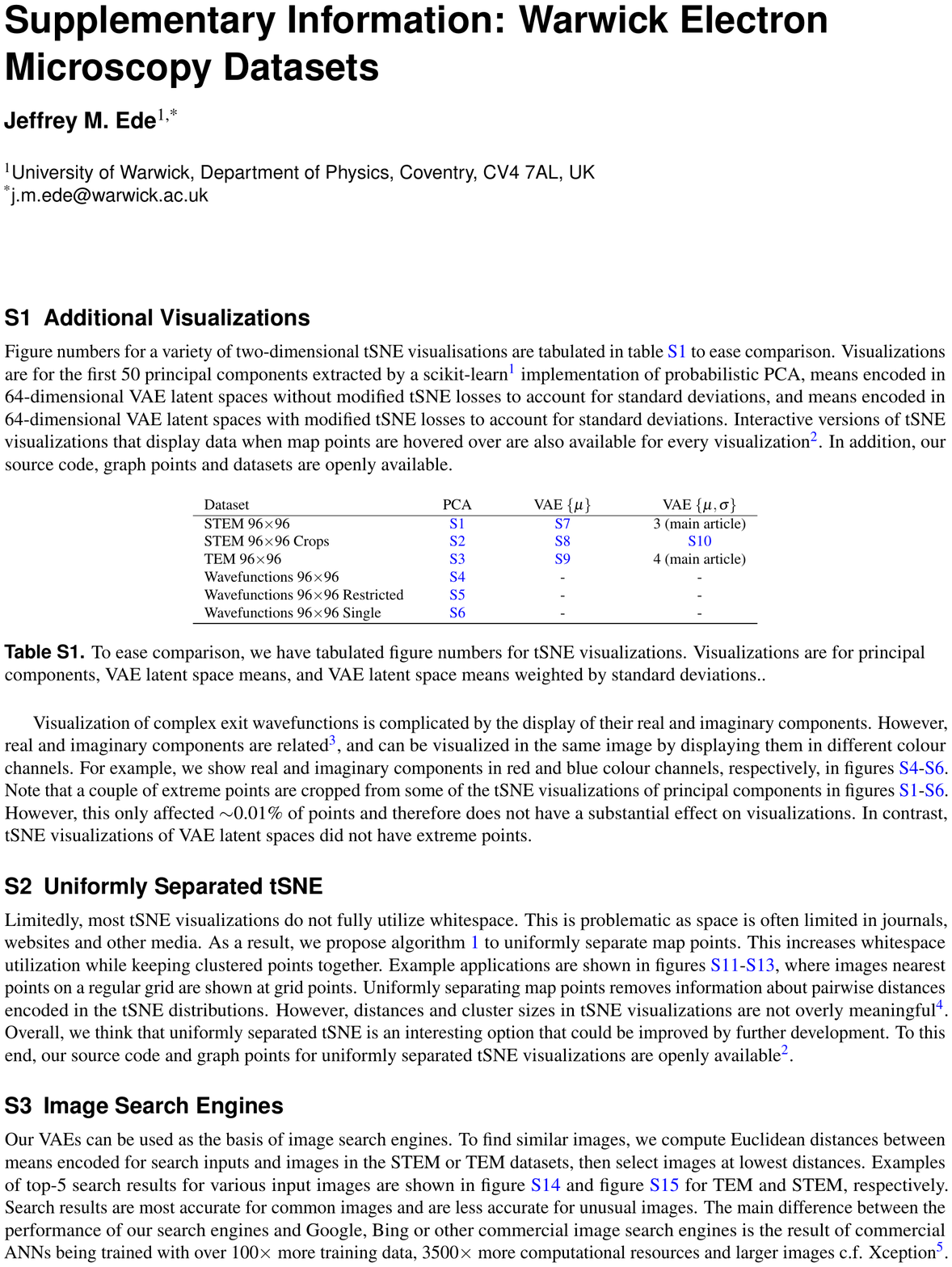}{%
  \ifthenelse{\value{imagepage}>0}{
  \newpage   
  \begingroup 
    \centering
    \includegraphics[
      trim={20mm 20.5mm 20mm 22.5mm},
      clip,
      page=\value{imagepage},
      width=\textwidth,  
      height=\textheight,
      keepaspectratio,
    ]{paper_wemd_supplementary}
    \newpage
  \endgroup
  }{}
}

\section{Amendments and Corrections}


There are amendments or corrections to the paper\cite{ede2020warwick} covered by this chapter. 

\correctionspacing
\noindent \textbf{Location:} Page~4, caption of fig.~2. \\
\textbf{Change:} \enquote{...at 500 randomly selected images...} should say \enquote{...at 500 randomly selected data points...}.

\section{Reflection}

This ancillary chapter covers my paper titled \enquote{Warwick Electron Microscopy Datasets}\cite{ede2020warwick} and associated research outputs\cite{ede2020warwick_supplementary, preprint+ede2020warwick, datasets_repo, warwickem}. My paper presents visualizations for large new electron microscopy datasets published with our earlier papers. There are 17266 TEM images curated to train our denoiser\cite{ede2019improving} (ch.~\ref{ch:denoiser}), 98340 STEM images curated to train generative adversarial networks (GANs) for compressed sensing\cite{ede2020partial, preprint+ede2019deep} (ch.~\ref{ch:partial_stem}), and 98340 TEM exit wavefunctions simulated to investigate EWR\cite{preprint+ede2020exit} (ch.~\ref{ch:ewr}), and derived datasets containing smaller TEM and STEM images that I created to rapidly prototype of ANNs for adaptive partial STEM\cite{preprint+ede2020adaptive_scans} (ch.~\ref{ch:adaptive_scans}). To improve visualizations, I developed new regularization mechanisms for variational autoencoders\cite{kingma2014auto, kingma2019introduction, doersch2016tutorial} (VAEs), which were trained to embed high-dimensional electron micrographs in low-dimensional latent spaces. In addition, I demonstrate that VAEs can be used as the basis of electron micrograph search engines. Finally, I provide extensions to t-distributed stochastic neighbour embedding\cite{maaten2008visualizing, linderman2019clustering, van2014accelerating, van2013barnes, wattenberg2016use} (tSNE) and interactive dataset visualizations.

Making our large machine learning datasets openly accessible enables our research to be reproduced\cite{baker2016reproducibility}, standardization of performance comparisons, and dataset reuse in future research. Dissemination of large datasets is enabled by the internet\cite{chinese2019development, berners1994uniform}, for example, through fibre optic\cite{kaushik2020fibre} broadband\cite{mack2020history, abrardi2019ultra} or satellite\cite{graydon2020connecting, kaufmanna2019performance} connections. Subsequently, there are millions of open access datasets\cite{castelvecchi2018google, noy2020discovering} that can be used for machine learning\cite{sun2017revisiting, hey2020machine}. Performance of ANNs usually increases with increasing training dataset size\cite{sun2017revisiting}, so some machine learning datasets have millions of examples. Examples of datasets with millions of examples include DeepMind Kinetics\cite{kay2017kinetics}, ImageNet\cite{russakovsky2015imagenet}, and YouTube 8M\cite{abu2016youtube}. Nevertheless, our datasets containing tens of thousands of examples are more than sufficient for initial exploration of deep learning in electron microscopy. For reference, some datasets used for initial explorations of deep learning for Coronavirus Disease 2019\cite{sohrabi2020world, wiersinga2020pathophysiology, singhal2020review} (COVID-19) diagnosis are 10$\times$ smaller\cite{teixeira2020impact} than WEMD.

There are many data clustering algorithms\cite{ghosal2020short, rodriguez2019clustering, djouzi2019review, mittal2019clustering, ansari2019spatiotemporal, jain1999data, lee2007nonlinear} that can group data for visualization. However, tSNE is a \textit{de facto} default as it often outperforms other algorithms\cite{maaten2008visualizing}. For context, tSNE is a variant of stochastic neighbour embedding\cite{hinton2003stochastic} (SNE) where a heavy-tailed Student's t-distribution is used to measure distances between embedded data points. Applications of tSNE include bioinformatics\cite{li2017application, wallach2009protein}, forensic science\cite{devassy2020dimensionality, melit2020unsupervised}, medical signal processing\cite{gang2018dimensionality, birjandtalab2016nonlinear, abdelmoula2016data}, particle physics\cite{psihas2019context, racah2016revealing}, smart electricity metering\cite{gong2020visual}, and sound synthesis\cite{mcdonald2018infinite}. Before tSNE, data is often embedded in a low-dimensional space to reduce computation, suppress noise, and prevent Euclidean distances used in tSNE optimization being afflicted by the curse of dimensionality\cite{schubert2017intrinsic}. For example, the original tSNE paper suggests using principal component analysis\cite{jolliffe2016principal, wall2003singular, halko2011finding, mart2011arand} (PCA) to reduce data dimensionality to 30 before applying tSNE\cite{maaten2008visualizing}.

Extensions of tSNE can improve clustering. For example, graphical processing unit accelerated implementations of tSNE\cite{pezzotti2019gpgpu, chan2018t} can speedup clustering 50-700$\times$. Alternatively, approximate tSNE\cite{pezzotti2016approximated} (A-tSNE) can trade accuracy for decreased computation time. Our tSNE visualizations took a couple of hours to optimize on an Intel i7-6700 central processing unit (CPU) as we used 10000 iterations to ensure that clusters stabilized. It follows that accelerated tSNE implementations may be preferable to reduce computation time. Another extension is to adjust distances used for tSNE optimization with a power transform based on the intrinsic dimension of each point. This can alleviate the curse of dimensionality for high-dimensional data\cite{schubert2017intrinsic}; however, it was not necessary for our data as I used VAEs to reduce image dimensionality to 64 before tSNE. Finally, tSNE early exaggeration (EE), where probabilities modelling distances in a high-dimensional space are increased, and number of iterations can be automatically tuned with opt-tSNE\cite{belkina2019automated}. Tuning can significantly improve visualizations, especially for large datasets with millions of examples. However, I doubt that opt-tSNE would result in large improvements to clustering as our datasets contain tens of thousands of examples, where tSNE is effective. Nevertheless, I expect that opt-tSNE could have improved clustering if I had been aware of it.

Further extensions to tSNE are proposed in my paper\cite{ede2020warwick, ede2020warwick_supplementary}. I think that the most useful extension uniformly separates clustered points based clustering density. Uniformly separated tSNE (US-tSNE) can often double whitespace utilization, which could make tSNE visualizations more suitable for journals,
websites, and other media where space is limited. However, the increased whitespace utilization comes at the cost of removing information about the structure of clusters. Further, my preliminary implementation of US-tSNE is limited insofar that Clough-Tocher cubic Bezier interpolation\cite{alfeld1984trivariate} used to map tSNE points to a uniform map is only applied to points within their convex hull. I also proposed a tSNE extension that uses standard deviations encoded by VAEs to inform clustering as this appeared to slightly improve clustering. However, I later found that using standard deviations appears to decrease similarity of nearest neighbours in tSNE visualizations. As a result, I think that how extra information encoded in standard deviations is used to inform clustering may merit further investigation.

To improve VAE encodings for tSNE, I applied a variant of batch normalization to their latent spaces. This avoids needing to tune a hyperparameter to balance VAE decoder and Kullback-Leibler (KL) losses, which is architecture-specific and can be complicated by relative sizes of their gradients varying throughout training. I also considered adaptive gradient balancing\cite{chen2017gradnorm, malkiel2019conditional} of losses; however, that would require separate backpropagation through the VAE generator for each loss, increasing computation. To increase image realism, I added Sobel losses to mean squared errors (MSEs). Sobel losses often improve realism as human vision is sensitive to edges\cite{mcilhagga2018estimates}. In addition, Sobel losses require less computation than VAE training with GAN\cite{larsen2016autoencoding} or perceptual\cite{grund2020improving} losses. Another computationally inexpensive approach to improve generated image realism is to train with structural similarity index measures\cite{wang2004image} (SSIMs) instead of MSEs\cite{zhao2016loss}.

My VAEs are used as the basis of my openly accessible electron microscopy search engines. I observe that top-5 search results are usually successful insofar that they contain images that are similar to input images. However, they often contain some images that are not similar, possibly due to there not being many similar images in our datasets. Thus, I expect that search results could be improved by increasing dataset size. Increasing input image size from 96$\times$96 to a couple of hundred pixels and increasing training iterations could also improve performance. Further, training could be modified to encode binary latent variables for efficient hashing\cite{dadaneh2020pairwise, pattersonsemantic, jin2019deep, mena2019binary, shen2018nash, chaidaroon2017variational}. Finally, I think that an interesting research direction is to create a web interface for an electron microscopy search engine that indexes institutional electron microscopy data servers. Such a search engine could enhance collaboration by making it easier to find electron microscopists working on interesting projects. 

An application of my VAEs that is omitted from my paper is that VAE generators could function as portable electron microscopy image generators. For example, to create training data for machine learning. For comparison, my VAE generators require roughly 0.1\% of the storage space needed for my image datasets to store their trainable parameters. However, I was concerned that a distribution of generated images might be biased by catastrophic forgetting\cite{liang2018generative}. Further, a distribution of generated images could be sensitive to ANN architecture and learning policy, including when training is stopped\cite{li2020gradient, flynn2020bounding}. Nevertheless, I expect that data generated from by VAEs could be used for pretraining to improve ANN robustness\cite{hendrycks2019using}. Overall, I think it will become increasingly practical to use VAEs or GANs as high-quality data generators as ANN architectures and learning policies are improved.

Perhaps the main limitation of my paper is that I did not introduce my preferred abbreviation, \enquote{WEMD}, for \enquote{Warwick Electron Microscopy Datasets}. Further, I did not define \enquote{WEMD} in my WEMD preprint\cite{preprint+ede2020warwick}. Subsequently, I introduced my preferred abbreviation in my review of deep learning in electron microscopy\cite{preprint+ede2020review} (ch.~\ref{ch:review}). I also defined an abbreviation, \enquote{WLEMD}, for \enquote{Warwick Large Electron Microscopy Datasets} in the first version of the partial STEM preprint\cite{preprint+ede2019partial} (ch.~\ref{ch:partial_stem}). Another limitation is that my paper only details datasets that had already been published, or that were derived from the published datasets. For example, Richard Beanland and I successfully co-authored an application for funding to simulate tens of thousands of CBED patterns with Felix\cite{beanland2020felix}, which are not detailed in my paper. The CBED dataset requires a couple of terabytes of storage and has not been processed for dissemination. Nevertheless, Richard Beanland\footnote{Email: r.beanland@warwick.ac.uk} may be able to provide the CBED dataset upon request.

\chapter{Adaptive Learning Rate Clipping Stabilizes Learning}\label{ch:ALRC}

\section{Scientific Paper}\label{sec:diss_adaptive}

This chapter covers the following paper\cite{ede2020adaptive}.
\begin{quote}
\bibentry{ede2020adaptive}
\end{quote}

\foreachpage{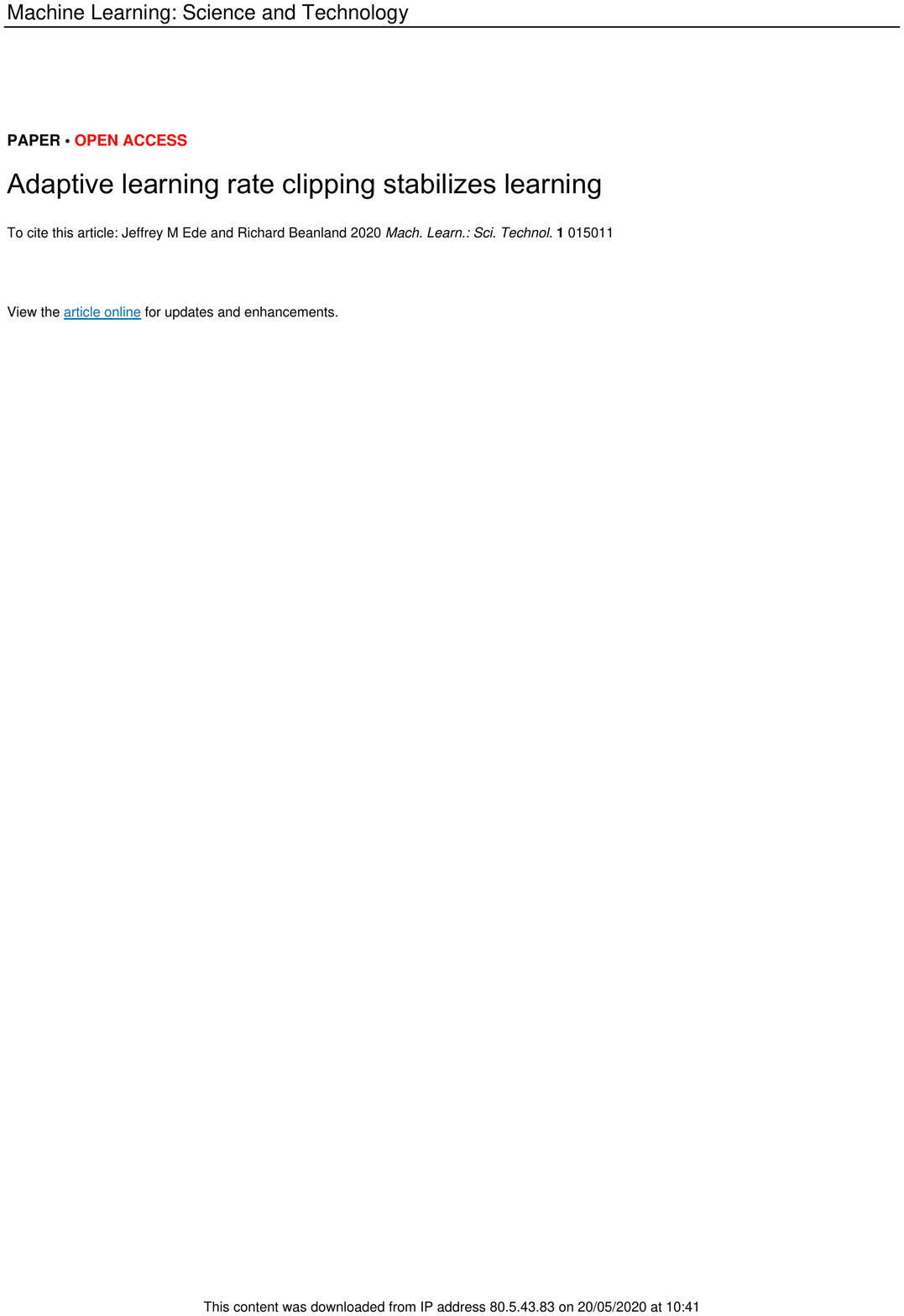}{%
  \ifthenelse{\value{imagepage}>1}{
  \newpage   
  \begingroup 
    \centering
    \includegraphics[
      trim={15.25mm 15.5mm 15.25mm 12.5mm},
      clip,
      page=\value{imagepage},
      width=\textwidth,  
      height=\textheight,
      keepaspectratio,
    ]{paper_alrc}
    \newpage
  \endgroup
  }{}
}

\section{Amendments and Corrections}

There are amendments or corrections to the paper\cite{ede2020adaptive} covered by this chapter. 

\correctionspacing
\noindent \textbf{Location:} Page~3, image in fig.~1. \\
\textbf{Change:} A title above the top two graphs is cut off. The missing title said \enquote{With Adaptive Learning Rate Clipping}, and is visible in our preprint\cite{preprint+ede2019adaptive}. 

\correctionspacing
\noindent \textbf{Location:} Last paragraph starting on page~7. \\
\textbf{Change:} \enquote{...inexpensive alternative to gradient clipping in high batch size training where...} should say \enquote{...inexpensive alternative to gradient clipping where...}.

\section{Reflection}

This ancillary chapter covers my paper titled \enquote{Adaptive Learning Rate Clipping Stabilizes Learning}\cite{ede2020adaptive} and associated research outputs\cite{preprint+ede2019adaptive, alrc_repo}. The ALRC algorithm was developed to prevent loss spikes destabilizing training of DNNs for partial STEM\cite{ede2020partial} (ch.~\ref{ch:partial_stem}). To fit the partial STEM ANN in GPU memory, it was trained with a batch size of 1. However, using a small batch size results in occasional loss spikes, which meant that it was sometimes necessary to repeat training to compare performance with earlier experiments where learning had not been destabilized by loss spikes. I expected that I could adjust training hyperparameters to stabilize learning; however, I had optimized the hyperparameters and training was usually fine. Thus, I developed ALRC to prevent loss spikes from destabilizing learning. Initially, ALRC was included as an appendix in the first version of the partial STEM preprint\cite{preprint+ede2019partial}. However, ALRC was so effective that I continued to investigate. Eventually, there were too many ALRC experiments to comfortably fit in an appendix of the partial STEM paper, so I separated ALRC into its own paper.

There are variety of alternatives to ALRC that can stabilize learning. A popular alternative is training with Huberized losses\cite{meyer2019alternative, huber1964robust},
\begin{equation}
    \text{Huber}(L) = \min(L, (\lambda L)^{1/2})\,,
\end{equation}
where $L$ is a loss and $\lambda$ is a training hyperparameter. However, I found that Huberized learning continued to be destabilized by loss spikes. I also considered gradient clipping\cite{seetharaman2020autoclip, pascanu2013difficulty, gorbunov2020stochastic}. However, my DNNs for partial STEM have many millions of trainable parameters, so computational requirements for gradient clipping are millions of times higher than applying ALRC to losses. Similarly, rectified ADAM\cite{liu2019variance} (RADAM), can stabilize learning by decreasing trainable parameter learning rates if adaptive learning rates of an ADAM\cite{kingma2014adam} optimizer have high variance. However, computational requirements of RADAM are also often millions of times higher than ALRC as RADAM adapts adaptive learning rates for every trainable parameter.

Overall, I think that ALRC merits further investigation. ALRC is computationally inexpensive, can be applied to any loss function, and appears to either stabilize learning or have no significant effect. Further, ALRC can often readily improve ANN training that would otherwise be destabilized loss spikes. However, I suspect that ALRC may slightly decrease performance where learning is not destabilized by loss spikes as ALRC modifies training losses. In addition, I have only investigated applications of ALRC to mean square and quartic errors per training example of deep convolutional neural networks (CNNs). Applying ALRC to losses for individual pixels of CNN outputs or to losses at each step of a recurrent neural network (RNN) may further improve performance. Encouragingly, my initial experiments with ALRC variants\cite{ede2020adaptive} show that a variety approaches improve training that would otherwise be destabilized by loss spikes.

\chapter{Partial Scanning Transmission Electron Microscopy with Deep Learning}\label{ch:partial_stem}

\section{Scientific Paper}\label{sec:diss_partial}

\noindent This chapter covers the following paper\cite{ede2020partial} and its supplementary information\cite{ede2020partial_supplementary}.
\begin{quote}
\bibentry{ede2020partial} 

\bibentry{ede2020partial_supplementary}
\end{quote}

 \foreachpage{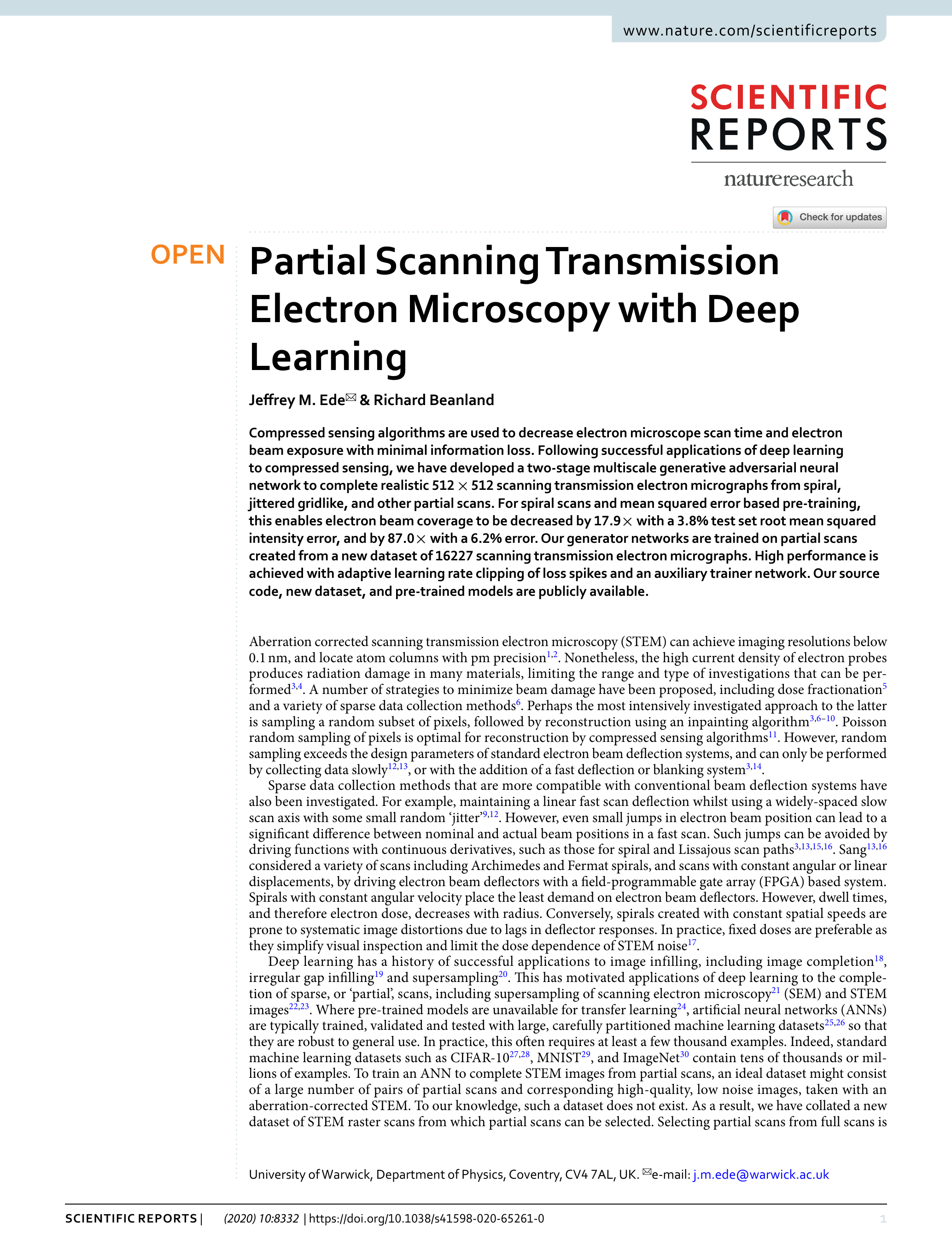}{%
  \ifthenelse{\value{imagepage}>0}{
  \newpage   
  \begingroup 
    \centering
    \includegraphics[
      trim={14.5mm 5.5mm 14.5mm 16.5mm},
      clip,
      page=\value{imagepage},
      width=\textwidth,  
      height=\textheight,
      keepaspectratio,
    ]{paper_partial_stem}
    \newpage
  \endgroup
  }{}
}

\foreachpage{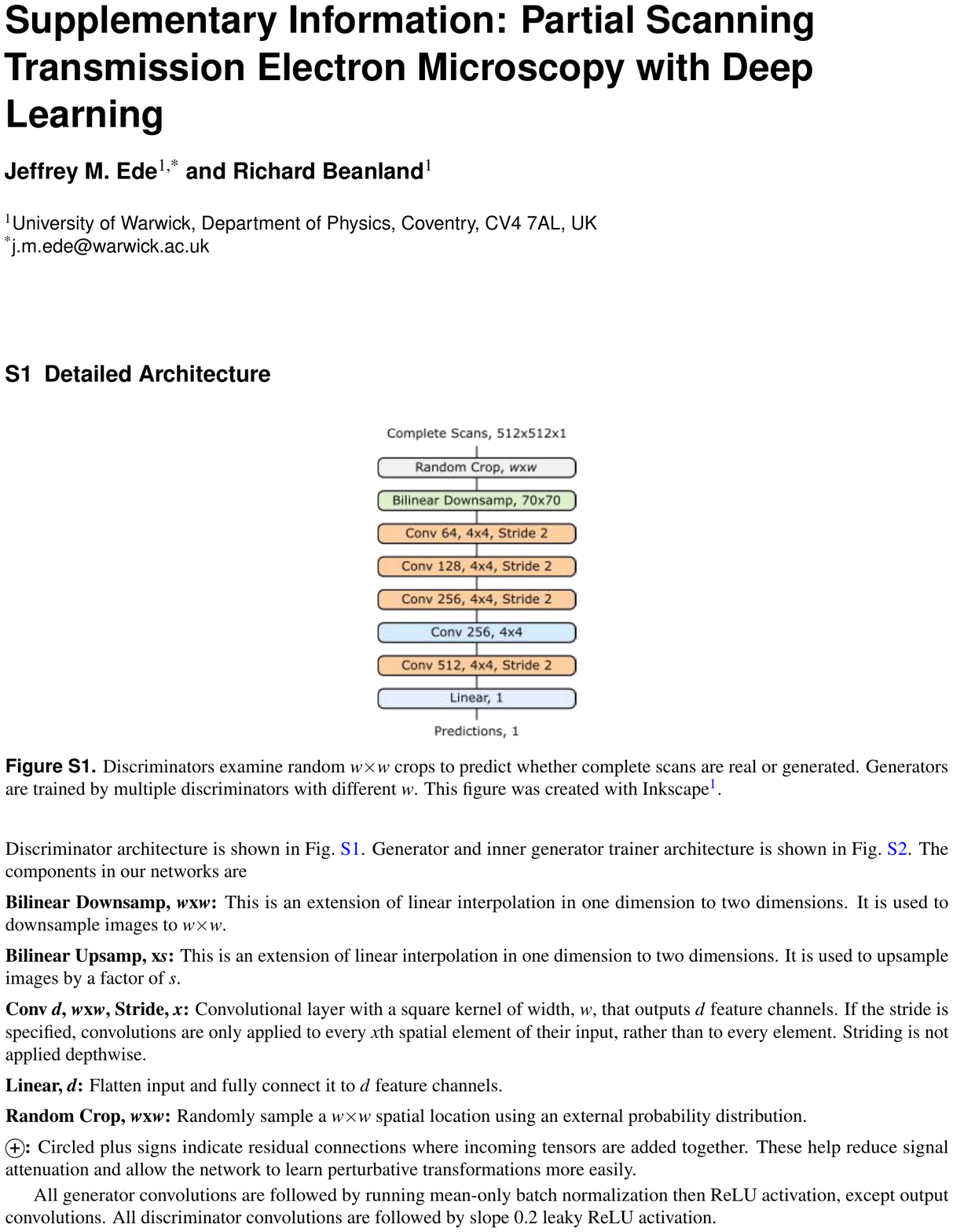}{%
  \ifthenelse{\value{imagepage}>0}{
  \newpage   
  \begingroup 
    \centering
    \includegraphics[
      trim={20mm 22mm 20mm 22.5mm},
      clip,
      page=\value{imagepage},
      width=\textwidth,  
      height=\textheight,
      keepaspectratio,
    ]{paper_partial_stem_supplementary}
    \newpage
  \endgroup
  }{}
}

\section{Amendments and Corrections}

There are amendments or corrections to the paper\cite{ede2020partial} covered by this chapter. 

\correctionspacing
\noindent \textbf{Location:} Reference 13 in the bibliography. \\
\textbf{Change:} \enquote{Sang, X. \textit{et al}. Dynamic Scan Control in STEM: Spiral Scans. Adv. Struct. Chem. Imaging 2, 6 (2017)} should say \enquote{Sang, X. \textit{et al}. Dynamic Scan Control in STEM: Spiral Scans. Adv. Struct. Chem. Imaging 2, 1–8 (2016)}.




\section{Reflection}

This chapter covers our paper titled \enquote{Partial Scanning Transmission Electron Microscopy with Deep Learning}\cite{ede2020partial} and associated research outputs\cite{ede2020partial_supplementary, preprint+ede2019partial, preprint+ede2019deep, unfinished+ede2020pixel, spirals_repo, dlss_repo, warwickem}, which were summarized by Bethany Connolly\cite{connolly2020atomic}. Our paper presents some of my investigations into compressed sensing of STEM images. Specifically, it combines results from two of my arXiv papers about compressed sensing with contiguous paths\cite{preprint+ede2019partial} and uniformly spaced grids\cite{preprint+ede2019deep} of probing locations. A third investigation into compressed sensing with a fixed random grid of probing locations was not published as I think that uniformly spaced grid scans are easier to implement on most scan systems. Further, reconstruction errors were usually similar for uniformly spaced and fixed random grids with the same coverage. Nevertheless, a paper I drafted on fixed random grids is openly accessible\cite{ede2020pixel}. Overall, I think that compressed sensing with DNNs is a promising approach to reduce electron beam damage and scan time by 10-100$\times$ with minimal information loss.

My comparison of spiral and uniformly spaced grid scans with the same ANN architecture, learning policy and training data indicates that errors are lower for uniformly spaced grids. However, the comparison is not conclusive as ANNs were trained for a few days, rather than until validation errors plateaued. Further, a fair comparison is difficult as suitability of architectures and learning policies may vary for different scan paths. Higher performance of uniformly spaced grids can be explained by content at the focus of most electron micrographs being imaged at 5-10$\times$ its Nyquist rate\cite{ede2020warwick} (ch.~\ref{ch:wemd}). It follows that high-frequency information that is accessible from neighbouring pixels in contiguous scans is often almost redundant. Overall, I think the best approach may combine both contiguous and uniform spaced grid scans. For example, a contiguous scan ANN could exploit high-frequency information to complete an image, which could then be mapped to a higher resolution image by an ANN for uniformly spaced scans. Indeed, functionality for contiguous and uniformly spaced grid scans could be combined into a single ANN. 

Most STEM scan systems can raster uniformly spaced grids of probing locations. However, scan systems often have to be modified to perform spiral or other custom scans\cite{sang2016dynamic, Sang2017a}. Modification is not difficult for skilled programmers. For example, Jonathan Peters\footnote{Email: petersjo@tcd.ie} created a custom scan controller prototype based on my field programmable gate array\cite{gandhare2019survey} (FPGA) within one day. Custom scans are often more distorted than raster scans. However, distortions can be minimized by careful choice of custom scan speed and path shape\cite{sang2016dynamic}. Alternatively, ANNs can correct electron microscope scan distortions\cite{zhang2017joint, jin2015correction}. We planned to use my FPGA to develop an openly accessible custom scan controller near the end of my PhD; however, progress was stalled by COVID-19 national lockdowns in the United Kingdom\cite{sinclair2020timeline}. As a result, I invested time that we had planned to use for FPGA deployment to review deep learning in electron microscopy\cite{preprint+ede2020review} (ch.~\ref{ch:review}).


To complete realistic images, generators were trained with MSEs or as part of GANs. However, GANs can introduce uncertainty into scientific investigation as they can generate realistic outputs, even if scan coverage is too low to reliably complete a region\cite{ede2020partial}. Consequently, investigated reducing uncertainty by adapting scan coverage\cite{preprint+ede2020adaptive_scans} to imaging regions (ch.~\ref{ch:adaptive_scans}). Alternatively, there are a variety of methods to quantify DNN uncertainty\cite{caldeira2020deeply, alaa2020uncertainty, staahl2020evaluation, loquercio2020general, kendall2019geometry, gal2016uncertainty, rudin2019stop}. For example, uncertainty can be predicted by ANNs\cite{van2020simple, lakshminarayanan2017simple}, Bayesian uncertainty approximation\cite{maddox2019simple, teye2018bayesian, kendall2017uncertainties, gal2016dropout}, or from variance of bootstrap aggregated\cite{breiman1996bagging} (bagged) model outputs. To address uncertainty, we present mean errors for 20000 test images, showing that errors are higher further away from scan paths. However, we do not provide an approach to quantify uncertainty of individual images, which could be critical to make scientific conclusions. Overall, I think that further investigation of uncertainty may be necessary before DNNs are integrated into default operating configurations of electron microscopes. 

A GAN could learn to generate any realistic STEM images, rather than outputs that correspond to inputs. To train GANs to generate outputs that correspond to inputs, I added MSEs between blurred input and output images to generator losses. Blurring prevented MSEs from strongly suppressing high-frequency noise characteristics. I also investigated adding distances between features output by discriminator layers for real and generated images to generator losses\cite{wang2018high}. However, feature distances require more computation than MSEs, and both feature distances and MSEs result in similar SSIMs\cite{ede2020pixel} between completed and true scans. As a result, I do not think that other computationally inexpensive additional losses, such as SSIMs or mean absolute errors, would substantially improve performance. Finally, I considered training generators to minimize perceptual losses\cite{santos2019learning}. However, most pretrained models used for feature extraction are not trained on electron micrographs or scientific images. Consequently, I was concerned that pretrained models might not clearly perceive characteristics specific to electron micrographs, such as noise.




\chapter{Adaptive Partial Scanning Transmission Electron Microscopy with Reinforcement Learning}\label{ch:adaptive_scans}

\section{Scientific Paper}\label{sec:diss_rl}

\noindent This chapter covers the following paper\cite{preprint+ede2020adaptive_scans} and its supplementary information\cite{ede2020adaptive_scans_supplementary}.
\begin{quote}
\bibentry{preprint+ede2020adaptive_scans}

\bibentry{ede2020adaptive_scans_supplementary}
\end{quote}

\foreachpage{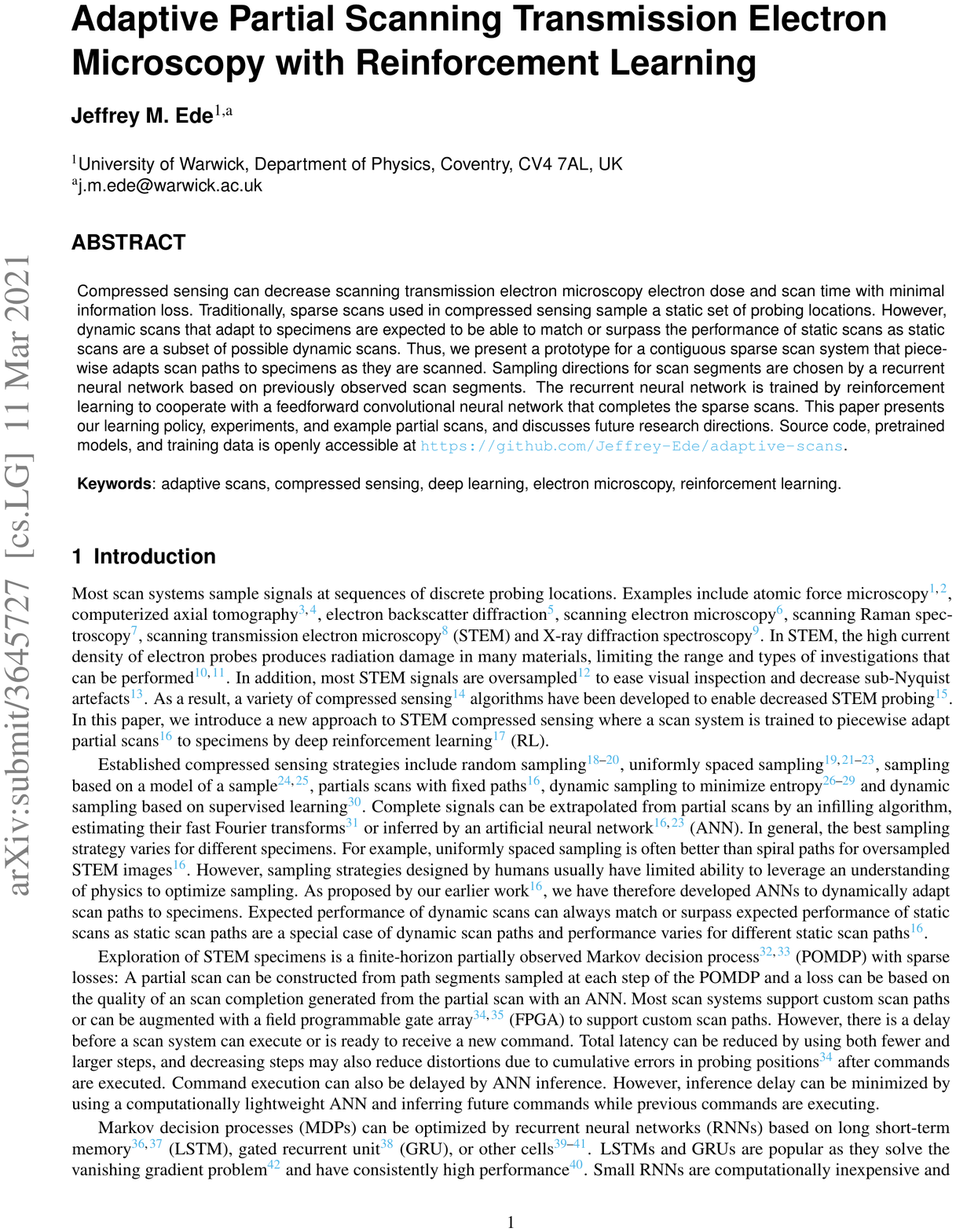}{%
  \ifthenelse{\value{imagepage}>0}{
  \newpage   
  \begingroup 
    \centering
    \includegraphics[
      trim={20mm 22mm 20mm 22.5mm},
      clip,
      page=\value{imagepage},
      width=\textwidth,  
      height=\textheight,
      keepaspectratio,
    ]{arxiv_adaptive_scans}
    \newpage
  \endgroup
  }{}
}

\foreachpage{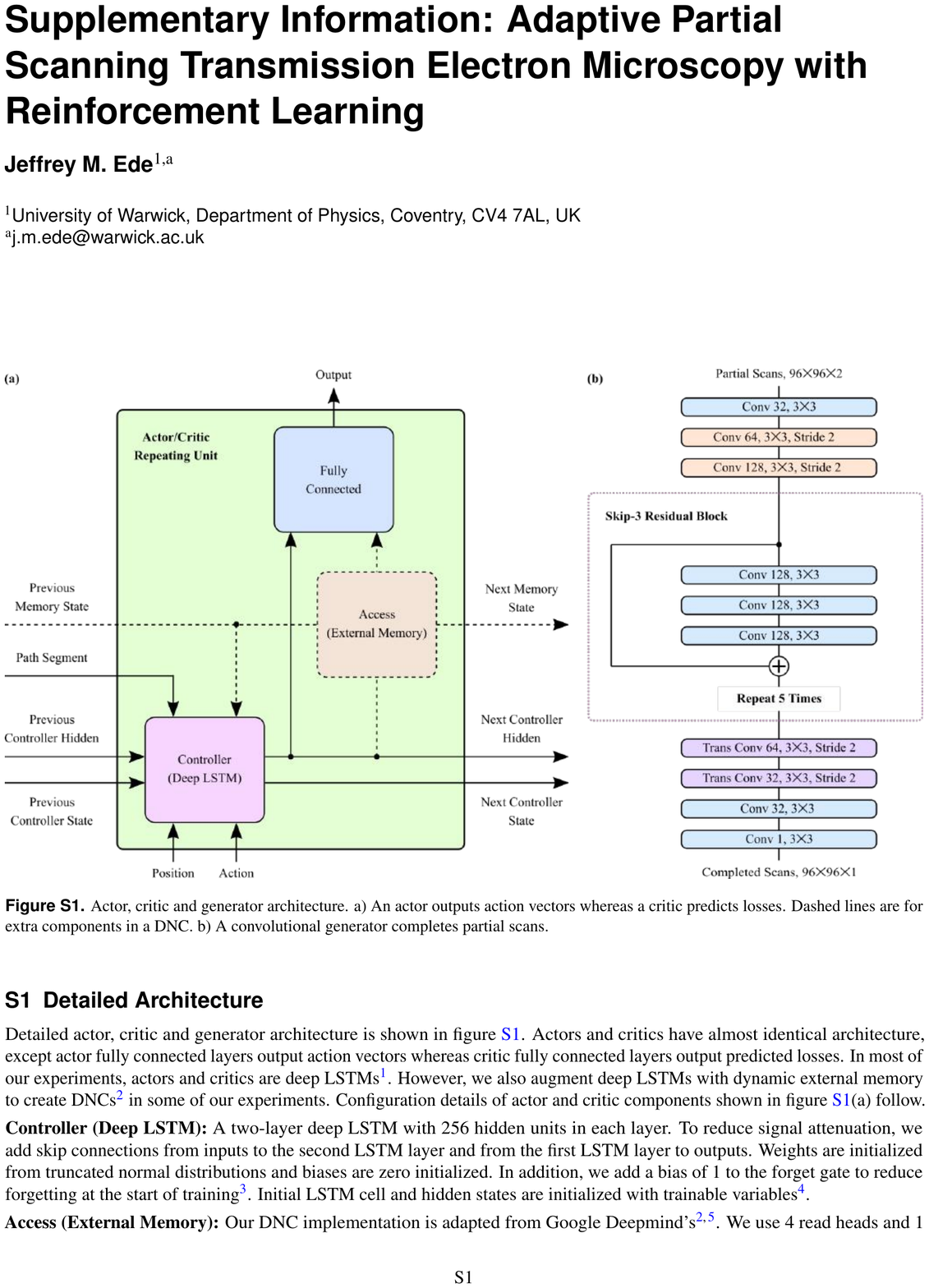}{%
  \ifthenelse{\value{imagepage}>0}{
  \newpage   
  \begingroup 
    \centering
    \includegraphics[
      trim={20mm 22mm 20mm 22.5mm},
      clip,
      page=\value{imagepage},
      width=\textwidth,  
      height=\textheight,
      keepaspectratio,
    ]{paper_adaptive_scans_supplementary}
    \newpage
  \endgroup
  }{}
}

\section{Reflection}

This chapter covers my paper titled \enquote{Adaptive Partial Scanning Transmission Electron Microscopy with Reinforcement Learning}\cite{preprint+ede2020adaptive_scans} and associated research outputs\cite{adaptive_scans_repo, warwickem}. It presents an initial investigation into STEM compressed sensing with contiguous scans that are piecewise adapted to specimens. Adaptive scanning is a finite-horizon partially observed Markov decision process\cite{saldi2019asymptotic, jaakkola1995reinforcement} (POMDP) with continuous actions and sparse rewards: Scan directions are chosen at each step based on previously observed path segments and a sparse reward is given by correctness completed sparse scans. Scan directions are decided by an actor RNN that cooperates with a generator CNN that completes full scans from sparse scans. Generator losses are not differentiable with respect to actor actions, so I introduced a differentiable critic RNN to predict generator losses from actor actions and observations. The actor and critic are trained by reinforcement learning with a new extension of recurrent deterministic policy gradients\cite{heess2015memory}, and the generator is trained by supervised learning.

This preliminary investigation was unsuccessful insofar that my prototype dynamic scan system does not convincingly outperform static scan systems. However, I believe that it is important to report my progress, despite publication bias against negative results\cite{earp2018need, mlinaric2017dealing, nissen2016publication, andrews2019identification, buvat2019dark, sharma2019positive, matosin2014negativity}, as it establishes starting points for further investigation. The main limitation of my scan system is that generator performance is much lower when it is trained for a variety of adaptive scan paths than when it is trained for a single static scan path. For an actor to learn an optimal policy, the generator should ideally be trained until convergence to the highest possible performance for every scan path. However, my generator architecture and learning policy was limited by available computational resources and development time. I also suspect that performance might be improved by replacing RNNs with transformers\cite{vaswani2017attention, alammar2018illustrated} as transformers often achieve similar or higher performance than RNNs\cite{karita2019comparative, zeyer2019comparison}.

There are a variety of additional refinements that could improve training. As an example, RNN computation is delayed by calling a Python function to observe each path segment. Delay could be reduced by more efficient sampling e.g. by using a parallelized routine coded in C/C++; by selecting several possible path segments in advance and selecting the segment that most closely corresponds to an action; or by choosing actions at least one step in advance rather than at each step. In addition, it may help if the generator undergoes additional training iterations in parallel to actor and critic training as improving the generator is critical to improving performance. Finally, increasing generator training iterations may result in overfitting, so it may help to train generators as part of a GAN or introduce other regularization mechanisms. For context, I find that adversarial training can reduce validation divergence\cite{preprint+ede2020exit} (ch.~\ref{ch:ewr}) and produce more realistic partial scan completions\cite{ede2020partial} (ch.~\ref{ch:partial_stem}).



\chapter{Improving Electron Micrograph Signal-to-Noise with an Atrous Convolutional Encoder-Decoder}\label{ch:denoiser}

\section{Scientific Paper}

\noindent This chapter covers the following paper\cite{ede2019improving}.
\begin{quote}
\bibentry{ede2019improving}
\end{quote}

\foreachpage{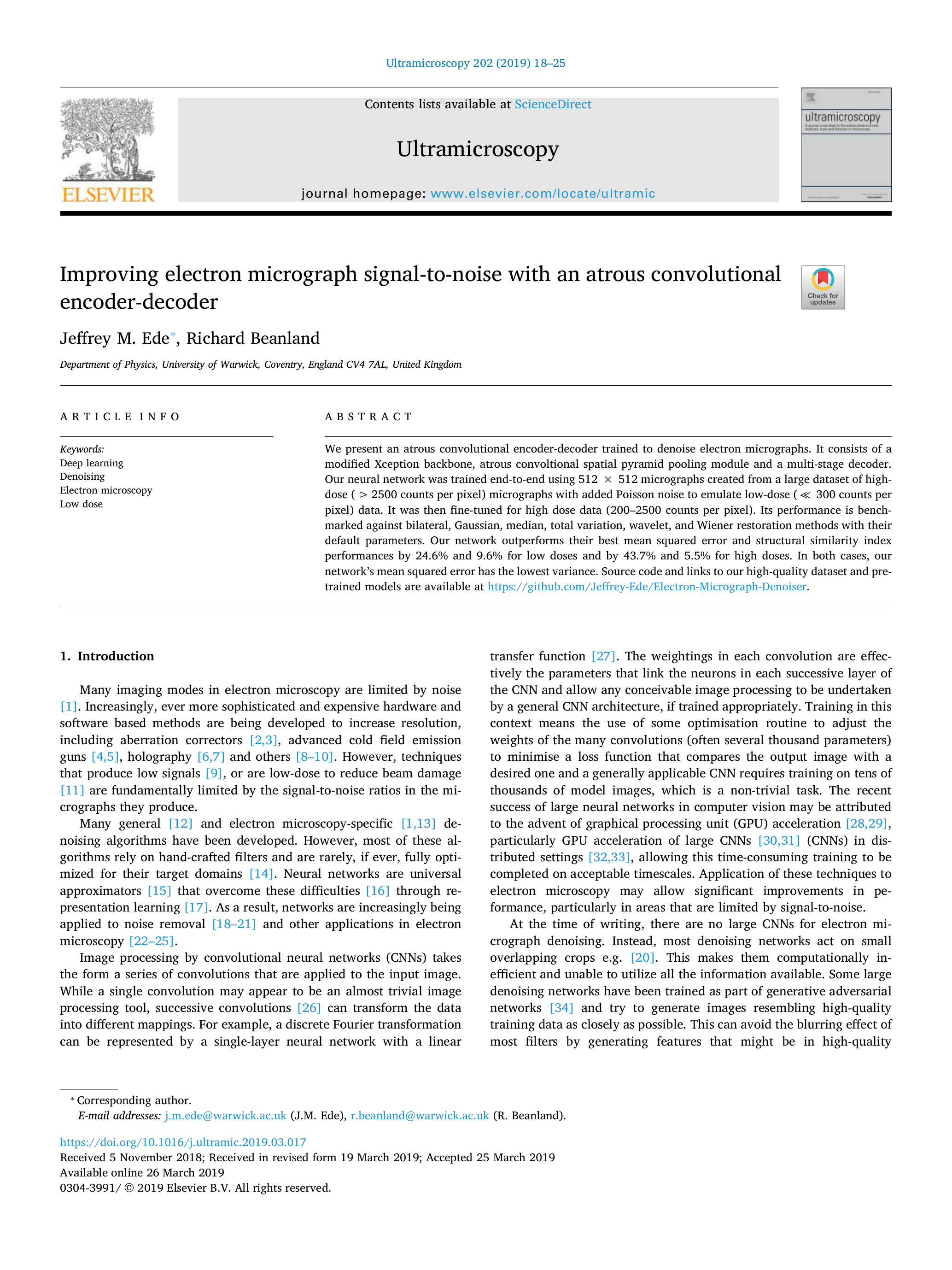}{%
  \newpage   
  \begingroup 
    \centering
    \includegraphics[
      trim={13mm 15.5mm 13mm 12.5mm},
      clip,
      page=\value{imagepage},
      width=\textwidth,  
      height=\textheight,
      keepaspectratio,
    ]{paper_denoiser}
    \newpage
  \endgroup
}

\section{Amendments and Corrections}

There are amendments or corrections to the paper\cite{ede2019improving} covered by this chapter. 

\correctionspacing
\noindent \textbf{Location:} Page~19, text following eqn~1. \\
\textbf{Change:} \enquote{...to only 25 $e^{-2}$ for a camera...} should say \enquote{...to only 25 $e \si{\angstrom}^{-2}$ for a camera...}.

\correctionspacing
\noindent \textbf{Location:} Page~21, first paragraph of performance section. \\
\textbf{Change:} \enquote{...structural similarity index (SSIM)...} should say \enquote{...structural similarity index measure (SSIM)...}.



\section{Reflection}

This chapter covers our paper titled \enquote{Improving Electron Micrograph Signal-to-Noise with an Atrous Convolutional Encoder-Decoder}\cite{ede2019improving} and associated research outputs\cite{preprint+ede2019improving, denoiser_repo, warwickem}. Our paper presents a DNN based on Deeplabv3+ that is trained to remove Poisson noise from TEM images. My DNN is affectionately named \enquote{Fluffles} and it is the only DNN that I have named. Pretrained models and performance characterizations are provided for DNNs trained for low and high electron doses. We also show that my DNN has lower MSEs, lower MSE variance, higher SSIMs, and lower or similar SSIM variance to other popular algorithms. We also provide MSE and SSIM distributions, and visualize errors for each output pixel. 

Due to limited available computational resources, DNN training was stopped after it surpassed the performance of a variety of popular denoising algorithms. However, there are many other denoising algorithms\cite{goyal2020image, girdher2019image, fan2019brief} that might achieve higher performance, some of which were developed for electron microscopy\cite{preprint+ede2020review}. For example, we did not compare our DNN against block-matching and 3D filtering\cite{dabov2007image, lebrun2012analysis} (BM3D), which often achieves high-performance. However, an extensive comparison is complicated by source code not being available for some algorithms. In addition, we expect that further training would improve performance as validation errors did not diverge from training errors. For comparison, our DNN was trained for about ten days on two Nvidia GTX 1080 Ti GPUs whereas Xception\cite{chollet2017xception}, which is randomly initialized as part of our DNN, was trained for one month on 60 Nvidia K80 GPUs for ImageNet\cite{krizhevsky2012imagenet} image classification. Indeed, I suspect that restarting DNN training with a pretrained Xception backbone may more quickly achieve much higher performance than continuing training from my pretrained models. Finally, sufficiently deep and wide ANNs are universal approximators\cite{kidger2019universal, lin2018resnet, hanin2017approximating, lu2017expressive, pinkus1999approximation, leshno1993multilayer, hornik1991approximation, hornik1989multilayer, cybenko1989approximation}, so denoising DNNs can always outperform or match the accuracy of other methods developed by humans.

A few aspects of my DNN architecture and optimization are peculiar as our paper presents some of my earliest experiments with deep learning. For example, learning rates were stepwise decayed at irregular \enquote{wall clock} times. Further, large decreases in errors when learning rates were decreased may indicate that learning rates were too high. Another issue is that ReLU6\cite{krizhevsky2010convolutional} activation does not significantly outperform ReLU\cite{nair2010rectified, glorot2011deep} activation, so ReLU is preferable as it requires less computation. Finally, I think that my DNN is too large for electron micrograph denoising. We justified that training can be continued and provide pretrained models; however, I doubt that training on the scale of Xception is practical insofar that most electron microscopists do not readily have access to more than a few GPUs for DNN training. I investigated smaller DNNs, which achieved lower performance. However, I expect that their performance could have been improved by further optimization of their training and architecture. In any case, I think that future DNNs for TEM denoising should be developed with automatic machine learning\cite{he2019automl, malekhosseini2019modeling, jaafra2019reinforcement, elsken2018neural, waring2020automated} (AutoML) as AutoML can balance accuracy and training time, and can often outperform human developers\cite{hanussek2020can, zoph2018learning}.

My denoiser has higher errors near output image edges. Higher errors near image edges were also observed for compressed sensing with spiral\cite{ede2020partial} and uniformly spaced grid\cite{preprint+ede2019deep} scans (ch.~\ref{ch:partial_stem}). Indeed, the structured systematic errors of my denoiser partially motivated my investigations of structured systematic errors in compressed sensing. To avoid higher errors at output edges, I overlap parts of images that my denoiser is applied to so that edges of outputs where errors are higher can be discarded. However, discarding parts of denoiser outputs is computationally inefficient. To reduce structured systematic errors, I tried weighting contributions of output pixel errors to training losses by multiplying pixel errors by their exponential moving averages\cite{ede2020partial}. However, weighting errors did not have a significant effect. Nevertheless, I expect that higher variation of pixel weights could reduce systematic errors. Moreover, I propose that weights for output pixel errors could be optimized during DNN training to minimize structured systematic errors.


\chapter{Exit Wavefunction Reconstruction from Single Transmission Electron Micrographs with Deep Learning}\label{ch:ewr}

\section{Scientific Paper}\label{sec:diss_exit}

\noindent This chapter covers the following paper\cite{preprint+ede2020exit} and its supplementary information\cite{ede2020wavefunctions_supplementary}.
\begin{quote}
\bibentry{preprint+ede2020exit}

\bibentry{ede2020wavefunctions_supplementary}
\end{quote}

\foreachpage{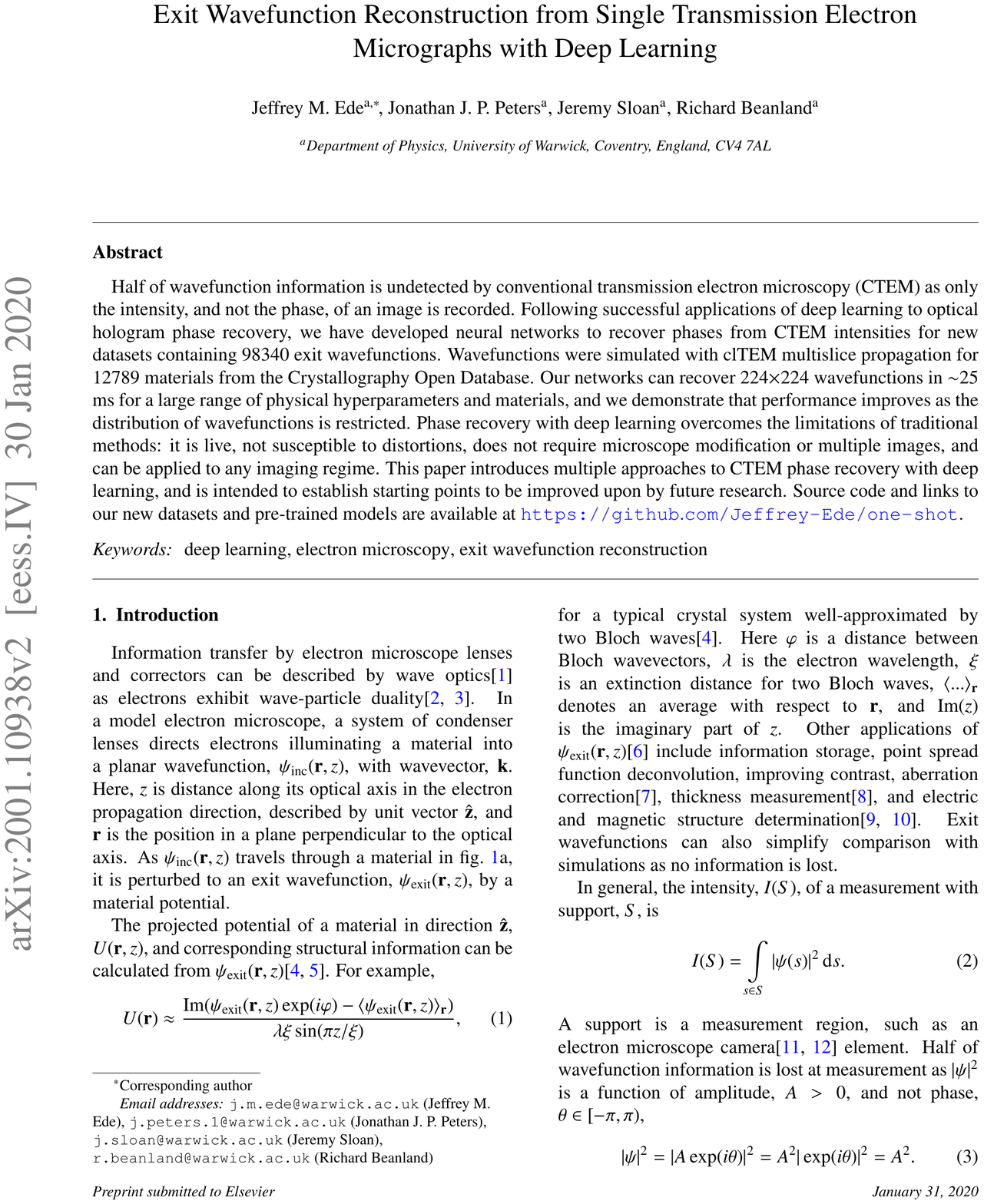}{%
  \ifthenelse{\value{imagepage}>0}{
  \newpage   
  \begingroup 
    \centering
    \includegraphics[
      trim={23mm 38.5mm 23mm 39mm},
      clip,
      page=\value{imagepage},
      width=\textwidth,  
      height=\textheight,
      keepaspectratio,
    ]{arxiv_wavefunctions}
    \newpage
  \endgroup
  }{}
}

\foreachpage{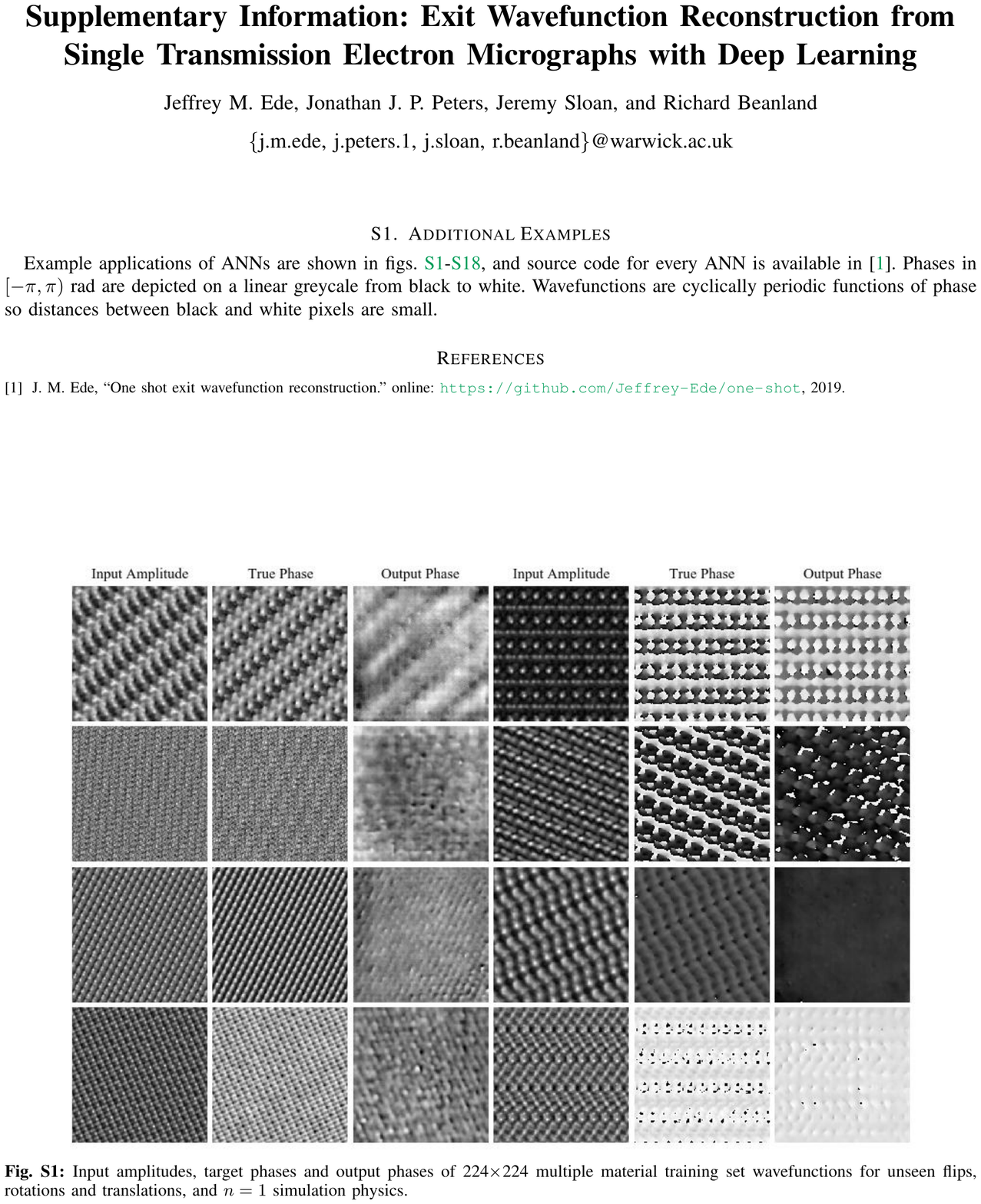}{%
  \ifthenelse{\value{imagepage}>0}{
  \newpage   
  \begingroup 
    \centering
    \includegraphics[
      trim={19mm 21.5mm 19mm 19mm},
      clip,
      page=\value{imagepage},
      width=\textwidth,  
      height=\textheight,
      keepaspectratio,
    ]{paper_wavefunctions_supplementary}
    \newpage
  \endgroup
  }{}
}

\section{Reflection}

This chapter covers our paper titled \enquote{Exit Wavefunction Reconstruction from Single Transmission Electron Micrographs with Deep Learning}\cite{preprint+ede2020exit} and associated research outputs\cite{one-shot_repo, warwickem}. At the University of Warwick, EWR is usually based on iterative focal and tilt series reconstruction (FTSR), so a previous PhD student, Mark Dyson, GPU-accelerated FTSR\cite{dyson2014advances}. However,  both recording a series of electron micrographs and FTSR usually take several seconds, so FTSR is unsuitable for live EWR. We have an electrostatic biprism that can be used for live in-line holography\cite{lehmann2002tutorial, koch2010off, ozsoy2014hybridization}; however, it is not used as we find that in-line holography is more difficult than FTSR. In addition, in-line holography can require expensive microscope modification if a microscope is not already equipped for it. Thus, I was inspired by applications of DNNs to predict missing information for low-light vision\cite{almasri2020robust, chen2018learning} to investigate live application of DNNs to predict missing phases of exit wavefunctions from single TEM images.

A couple of years ago, it was shown that DNNs can recover phases of exit wavefunctions from single optical micrographs if wavefunctions are constrained by limiting input variety\cite{rivenson2018phase, wu2018extended, sinha2017lensless}. Similarly, electron propagation can be described by wave optics\cite{rose2008optics}, and optical and electron microscopes have similar arrangements of optical and electromagnetic lenses, respectively\cite{chen2011optical}. Thus, it might be expected that DNNs can recover phases of exit wavefunctions from single TEM images. However, earlier experiments with optical micrographs were unbeknownst to us when we started our investigation. Thus, whether DNNs could reconstruct phase information from single TEM images was contentions as there are infinite possible phases for a given amplitude. Further, previous non-iterative approaches to TEM EWR were limited to defocused images in the Fresnel regime\cite{morgan2011direct} or non-planar incident wavefunctions in the Fraunhofer regime\cite{martin2008direct}.

We were not aware of any large openly accessible datasets containing experimental TEM exit wavefunctions. Consequently, we simulated exit wavefunctions with clTEM\cite{clTEM_repo, dyson2014advances} for a preliminary investigation. Similar to optical EWR\cite{rivenson2018phase, wu2018extended, sinha2017lensless}, we found that DNNs can recover the phases of TEM exit wavefunctions if wavefunction variety is restricted. Limitingly, our simulations are unrealistic insofar they do not include aberrations, specimen drift, statistical noise, and higher-order simulation physics. However, we have demonstrated that DNNs can learn to remove noise\cite{ede2019improving} (ch.~\ref{ch:denoiser}), specimen drifted can be reduced by sample holders\cite{goodge2020atomic}, and aberrations can be minimized by aberration correctors\cite{rose2008optics, pennycook2017impact, ramasse2017twenty, hawkes2009aberration}. Moreover, our results present lower bounds for performance as our inputs were far less restricted than possible in practice. 

Curating a dataset of experimental exit wavefunctions to train DNNs to recover their phases is time-consuming and expensive. Further, data curation became impractical due to a COVID-19 national lockdown in the United Kingdom\cite{sinclair2020timeline}. Instead, we propose a new approach to EWR that uses metadata to inform DNN training with single images. Our TEM (ch.~\ref{ch:denoiser}) and STEM (ch.~\ref{ch:partial_stem}) images in WEMD\cite{ede2020warwick} are provided as a possible resource to investigate our proposal. However, metadata is not included in WEMD, which is problematic as performance is expected to increase with increasing metadata as increasing metadata increasingly restricts probable exit wavefunctions. Nevertheless, DNNs can reconstruct some metadata from unlabelled electron micrographs\cite{weber2018automated}. Another issue is that experimental WEMD contain images for a range of electron microscope configurations, which would complicate DNN training. For example, experimental TEM images include bright field, dark field, diffraction and CBED images. However, data clustering could be applied to partially automate labelling of electron microscope configurations. For example, I provide pretrained VAEs to embed images for tSNE\cite{ede2020warwick} (ch.~\ref{ch:wemd}).



\chapter{Conclusions}\label{ch:conclusions}

This thesis covers a subset of my papers on advances in electron microscopy with deep learning. My review paper (ch.~\ref{ch:review}) offers a substantial introduction that sets my work in context. Ancillary chapters then introduce new machine learning datasets for electron microscopy (ch.~\ref{ch:wemd}) and an algorithm to prevent learning instabilty when training large neural networks with limited computational resources (ch.~\ref{ch:ALRC}). Finally, we report applications of deep learning to compressed sensing in STEM with static (ch.~\ref{ch:partial_stem}) and dynamic (ch.~\ref{ch:adaptive_scans}) scans, improving TEM signal-to-noise (ch.~\ref{ch:denoiser}), and TEM exit wavefunction reconstruction (ch.~\ref{ch:ewr}). This thesis therefore presents a substantial original contribution to knowledge which is, in practice, worthy of peer-reviewed publication. This thesis adds to my existing papers by presenting their relationships, reflections, and holistic conclusions. To encourage further investigation, source code, pretrained models, datasets, and other research outputs associated with this thesis are openly accessible. 

Experiments presented in this thesis are based on unlabelled electron microscopy image data. Thus, this thesis demonstrates that large machine learning datasets can be valuable without needing to add enhancements, such as image-level or pixel-level labels, to data. Indeed, this thesis can be characterized as an investigation into applications of large unlabelled electron microscopy datasets. However, I expect that tSNE clustering based on my pretrained VAE encodings\cite{ede2020warwick} (ch.~\ref{ch:wemd}) could ease image-level labelling for future investigations. Most areas of science are facing a reproducibility crisis\cite{baker2016reproducibility}, including artificial intelligence\cite{hutson2018artificial}, which I think is partly due to a perceived lack of value in archiving data that has not been enhanced. However, this thesis demonstrates that unlabelled data can readily enable new applications of deep learning in electron microscopy. Thus, I hope that my research will encourage more extensive data archiving by the electron microscopy community.

My DNNs were developed with TensorFlow\cite{abadi2016tensorflow, abadi2016tensorflow_large} and Python. In addition, recent versions of Gatan Microscopy Suite (GMS) software\cite{gms_webpage}, which is often used to drive electron microscopes, support Python\cite{miller2019real}. Thus, my pretrained models and source code can be readily integrated into existing GMS software. If a microscope is operated by alternative software or an older version of GMS that does not support Python, TensorFlow supports many other programming languages\cite{preprint+ede2020review} which can also interface with my pretrained models, and which may be more readily integrated. Alternatively, Python code can often be readily embedded in or executed by other programming languages. To be clear, my DNNs were developed as part of an initial investigation of deep learning in electron microscopy. Thus, this thesis presents lower bounds for performance that may be improved upon by refining ANN architecture and learning policy. Nevertheless, my pretrained models can be the initial basis of deep learning software for electron microscopy.

This thesis includes a variety of experiments to refine ANN architecture and learning policy. As AutoML\cite{he2019automl, malekhosseini2019modeling, jaafra2019reinforcement, elsken2018neural, waring2020automated} has improved since the start of my PhD, I expect that human involvement can be reduced in future investigations of standard architecture and learning policy variations. However, AutoML is yet to be able to routinely develop new approaches to machine learning, such as VAE encoding normalization and regularization\cite{ede2020warwick} (ch.~\ref{ch:wemd}) and ALRC\cite{ede2020adaptive} (ch.~\ref{ch:ALRC}). Most machine learning experts do not think that a technological singularity, where machines outrightly surpasses human developers, is likely for at least a couple of decades\cite{muller2016future}. Nonetheless, our increasingly creative machines are already automating some aspects of software development\cite{sarkar2020towards, guzdial2018co} and can programmatically describe ANNs\cite{sethi2017dlpaper2code}. Subsequently, I encourage adoption of creative software, like AutoML, to ease development.

Perhaps the most exciting aspect of ANNs is their scalability\cite{lwakatare2020large, gupta2016scalable}. Once an ANN has been trained, clones of the ANN and supporting software can be deployed on many electron microscopes at little or no additional cost to the developer. All machine learning software comes with technical debt\cite{breck2017ml, sculley2015hidden}; however, software maintenance costs are usually far lower than the cost of electron microscopes. Thus, machine learning may be a promising means to cheaply enhance electron microscopes. As an example, my experiments indicate that compressed sensing ANNs\cite{ede2020partial} (ch.~\ref{ch:partial_stem}) can increase STEM and other electron microscopy resolution by up to 10$\times$ with minimal information loss. Such a resolution increase could greatly reduce the cost of electron microscopes while maintaining similar capability. Further, I anticipate that multiple ANNs offering a variety of functionality can be combined into a single- or multiple-ANN system that simultaneously offers a variety of enhancements, including increased resolution, decreased noise\cite{ede2019improving} (ch.~\ref{ch:denoiser}), and phase information\cite{preprint+ede2020exit} (ch.~\ref{ch:denoiser}).

I think the main limitation of this thesis, and deep learning, is that it is difficult to fairly compare different approaches to DNN development. As an example, I found that STEM compressed sensing with regularly spaced scans outperforms contiguous scans for the same ANN architecture and learning policy\cite{ede2020partial} (ch.~\ref{ch:partial_stem}). However, such a performance comparison is complicated by sensitivity of performance to training data, architecture, and learning policy. As a case in point, I argued that contiguous scans could outperform spiral scans if STEM images were not oversampled\cite{ede2020partial}, which could be the case if partial STEM ANNs are also trained to increase image resolution. In part, I think ANN development is an art: Most ANN architecture and learning policy is guided by heuristics, and best approaches to maximize performance are chosen by natural selection\cite{johnson2010self}. Due to the complicated nature of most data, maximum performances that can be achieved with deep learning are not known. However, it follows from the universal approximator theorem\cite{kidger2019universal, lin2018resnet, hanin2017approximating, lu2017expressive, pinkus1999approximation, leshno1993multilayer, hornik1991approximation, hornik1989multilayer, cybenko1989approximation} that minimum errors can, in principle, be achieved by DNNs.

Applying an ANN to a full image usually requires less computation than applying an ANN to multiple image crops. Processing full images avoids repeated calculations if crops overlap\cite{ede2019improving} (ch.~\ref{ch:denoiser}) or lower performance near crop edges where there is less information\cite{ede2019improving, ede2020partial, preprint+ede2019deep} (ch.~\ref{ch:partial_stem} and ch.~\ref{ch:denoiser}). However, it is usually impractical to train large DNNs to process full electron microscopy images, which are often 1024$\times$1024 or larger, due to limited memory in most GPUs. This was problematic as one of my original agreements about my research was that I would demonstrate that DNNs could be applied to large electron microscopy images, which Richard Beanland and I decided were at least 512$\times$512. As a result, most of my DNNs were developed for 512$\times$512 crops from electron micrographs, especially near the start of my PhD. The combination of large input images and limited available GPU memory restricted training batch sizes to few examples for large ANNs, so I often trained ANNs with a batch size of 1 and either weight\cite{salimans2016weight} or spectral\cite{miyato2018spectral} normalization, rather than batch normalization\cite{ioffe2015batch}.

Most of my DNNs leverage an understanding of physics to add extra information to electron microscopy images. Overt examples include predicting unknown pixels for compressed sensing with static\cite{ede2020partial} (ch.~\ref{ch:partial_stem}) or adaptive\cite{preprint+ede2020adaptive_scans} (ch.~\ref{ch:adaptive_scans}) sparse scans, and unknown phase information from image intensities\cite{preprint+ede2020exit} (ch.~\ref{ch:ewr}). More subtly, improving image signal-to-noise with an DNN\cite{ede2019improving} (ch.~\ref{ch:denoiser}) is akin to improving signal-to-noise by increasing numbers of intensity measurements. Arguably, even search engines based on VAEs\cite{ede2020warwick} (ch.~\ref{ch:wemd}) add information to images insofar that VAE encodings can be compared to quantify semantic similarities between images. Ultimately, my DNNs add information to data that could already be understood from physical laws and observations. However, high-dimensional datasets can be difficult to utilize. Deep learning offers an effective and timely means to both understand high-dimensional data and leverage that understanding to produce results in a useable format. Thus, I both anticipate and encourage further investigation of deep learning in electron microscopy.


\clearpage
\renewcommand{\bibname}{References}
\bibliography{bibliography.bib}            

\begin{thesisauthorvita}             

\noindent This vita covers the following resume\cite{ede2020redacted}.

 \begin{quote}
 \bibentry{ede2020redacted}
 \end{quote}

\foreachpage{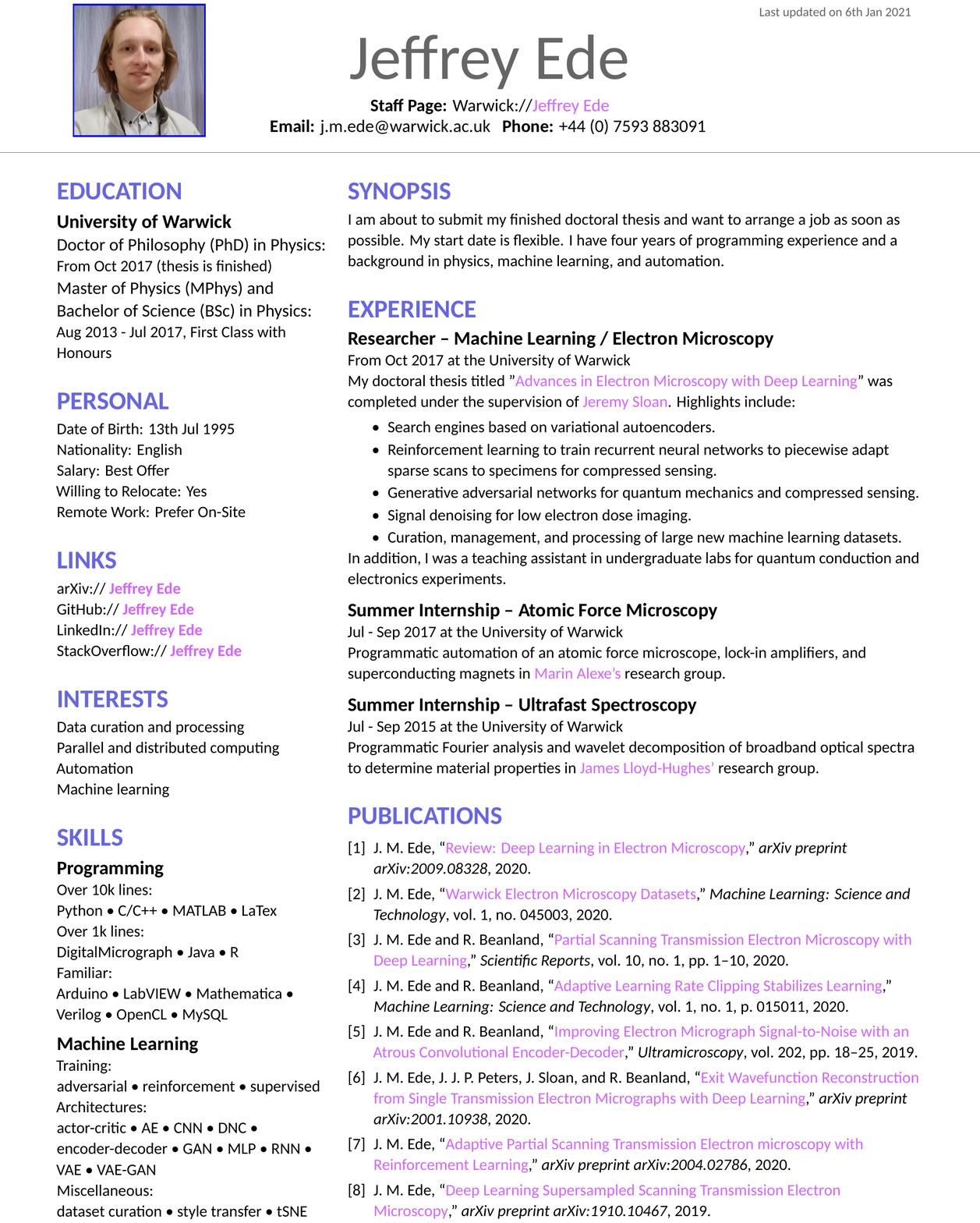}{%
  \ifthenelse{\value{imagepage}>0}{
  \newpage   
  \begingroup 
    \centering
    \includegraphics[
      trim={12.5mm 0.75mm 12.5mm 0.75mm},
      clip,
      page=\value{imagepage},
      width=\textwidth,  
      height=\textheight,
      keepaspectratio,
    ]{resume}
    \newpage
  \endgroup
  }{}
}



 


\end{thesisauthorvita}              

\end{document}